\documentclass[prl,aps,twocolumn,groupaddress]{revtex4-1}
\usepackage{latexsym}
\usepackage{hyperref}
\usepackage{url}
\usepackage{graphicx}
\usepackage{verbatim}
\usepackage{multirow}
\usepackage{amsmath}
\newcommand{\beq}{\begin{eqnarray}}
\newcommand{\eeq}{\end{eqnarray}}
\usepackage{mathrsfs}
\usepackage{float}
\usepackage[usenames, dvipsnames]{color}
\usepackage{mathtools}
\usepackage{slashed}
\usepackage{physics}	
\usepackage{graphicx}   
\usepackage{epstopdf}
\usepackage{subfigure}  
\usepackage{hyperref}   
\usepackage{bbold}
\usepackage{wasysym}
\usepackage{feynmp}
\usepackage[symbol]{footmisc}
\def \bs{\textbf}
\usepackage[latin1]{inputenc}
\usepackage{tikz}
\usetikzlibrary{decorations.pathmorphing}
\usetikzlibrary{shapes.misc}
\tikzset{cross/.style={cross out, draw=black, minimum size=8*(#1-\pgflinewidth), inner sep=0pt, outer sep=0pt},
cross/.default={1pt}}
\usetikzlibrary{patterns,math}
\begin{document}

\title{Pairing instability on a Luttinger surface: A non-Fermi liquid to
    superconductor transition and its Sachdev-Ye-Kitaev dual }
\author{Chandan Setty}
\thanks{email for correspondence: csetty@ufl.edu}
\affiliation{Department of Physics, University of Florida, Gainesville, Florida, USA}

\begin{abstract}
\textcolor{black}{Superconductivity results from an instability of the Fermi surface -- contour of \textit{poles} of the single particle propagator -- to an infinitesimally small attraction between electrons. Here, we instead discuss the analogous problem on a model \textit{Luttinger} surface, or contour of \textit{zeros} of the Green function.} At zero temperature ($\beta \rightarrow \infty$) and a critical interaction strength ($u_{c\infty}$) characterized by the residue of self-energy pole, we find that the pair susceptibility diverges leading to a superconducting instability.  We evaluate the pair fluctuation partition function and find that the spectral density in the normal state has an interaction-driven, power-law $\frac{1}{\sqrt{\omega}}$ type, van-Hove singularity (vHS) indicating non-Fermi liquid (NFL) physics. \textcolor{black}{Crucially, in the strong coupling limit ($\beta u \gg 1$), the leading order fluctuation free energy terms in the normal state of this NFL-SC transition resembles the equivalent $\left(O(1)\right)$ terms of the Sachdev-Ye-Kitaev (SYK) model. This free energy contribution takes a simple form $-\beta F = \beta u_{c\infty} - \gamma~\text{ln}\left(\beta u_{c \infty}\right)$ where $\gamma$ is a constant equal to $\frac{1}{2}$.} 
Weak impurity scattering ($\tau \gg \beta^{-1}$) leaves the low-energy spectral density unaffected, but leads to an interaction-driven enhancement of superconductivity.  Our results 
 shed light on the role played by order-parameter fluctuations in providing the key missing link between Mott physics and strongly coupled 
toy-models exhibiting gravity duals.
\end{abstract}

\maketitle
\section{Introduction} 
A central notion that captures the failure of single-particle physics in quantum matter is the Luttinger surface (LS) -- a contour in momentum space where the many-body Green function, $G(\bs p, \omega)$, vanishes~\cite{AGD1965}. This lies in contrast to the normal Fermi Liquid (FL) where  particle excitations are characterized by poles in the single-particle propagator. The LS has been invoked to reconcile several key experimental observations~\cite{Statt1999, Uchida2006, Shen2010,Shen2011, Kidd2011, Chakravarty2010} in the Cuprate-Mott insulator under a single unifying paradigm, including the Luttinger sum rule (LSR)~\cite{Luttinger1960, Ward1960, Oshikawa2000} and its apparent violation ~\cite{ Dzyaloshinskii1996, Dzyaloshinskii2003, Tsvelik2002, Tsvelik2006, Zhang2006, Morr1998, Kotliar2006, Choy2007, Rosch2007, Kane2013, Georges2006, Imada2009}, pseudo-gap and Fermi arcs~\cite{Zhang2006, Kotliar2006, Phillips2006, Bascones2007, Campuzano2007, Vanacore2018}, spectral weight transfer~\cite{Phillips2006, Choy2007}, as well as features in the self-energy, $\Sigma(\bs p, \omega)$~\cite{Sachdev2018}. \par
A salient property of the LS which gives rise to the aforementioned observations is a divergent $\Sigma(\bs p, \omega)$~\cite{AGD1965, Tsvelik2002, Tsvelik2006, Zhang2006, Kotliar2006, Hong-Phillips2012, Sachdev2018}. The breakdown of the LSR -- a rule which relates the density of electrons at fixed chemical potential to the number of excitations in the FL  and whose generalizations were shown to hold in broader contexts~\cite{Bedell1997, Affleck1997,Oshikawa2000,  Tsvelik2002, Dzyaloshinskii2003, Yunoki2017, Bedell2019} -- 
 serves as an illustrative example to highlight the consequences of a singular self-energy. While the total particle density equals the area enclosed by the surface of propagator-poles when $\Sigma(\bs p, \omega)$ is regular, there is an anomalous contribution to the density, proportional to $I = \int G \frac{\partial \Sigma}{\partial \omega}$, that averages to zero in a FL~\cite{Luttinger1960, Ward1960, AGD1965}.  The integral $I$ counts the excess density in addition to the volume contained inside contours where $G(\bs p, \omega)$ changes sign~\cite{AGD1965, Dzyaloshinskii1996, Dzyaloshinskii2003} and can, however, be non-vanishing when $\Sigma(\bs p, \omega)$ diverges~\cite{Sachdev2001, Kotliar2006, Kane2013}. These many-body properties follow entirely from explicit electron-electron interactions in the problem. \par
Nevertheless, the normal state of a superconductor can exhibit anomalies that deviate from a FL even in the absence of explicit electron correlations. This class of phenomena originates from Cooper-pair fluctuations~\cite{LarkinVarlamov2005, Glatz2018} and lead to precursor effects  wherein certain characteristics of the SC are retained for temperatures $T>T_c$, and in some cases, can even persist for $T\gg T_c$. With knowledge of the fluctuation propagator $L(\bs q, \omega)$ -- the fundamental object in the theory of pair fluctuations constructed from the ground state of the system for $T>T_c$ -- various measurable quantities can be evaluated systematically and compared with experiment~\cite{LarkinVarlamov2005, Glatz2018}. Several observations such as paraconductivity, rounding of transverse resistance peak, excess tunneling current, pseudo-gap behavior etc (see Refs.~\cite{Varlamov1991, Varlamov1997},~\cite{Varlamov1993},~\cite{Goldman1972}, for example, as well as ~\cite{Glatz2018, LarkinVarlamov2005} for a more detailed review) have been successfully understood via fluctuation physics derived from a free electron Green function. More generic models describing the thermodynamics of fluctuations in multi-band systems have also been examined in the context of MgB$_2$~\cite{Vinokur2005}. \par
In this work, we introduce interactions explicitly by \textcolor{black}{analyzing pairing instability and pair fluctuations on a system with a LS formed by a pole in $\Sigma$.  Unlike the problem of pairing on a Fermi surface,} we find a quantum phase transition into the superconducting state at a critical interaction strength ($u_{c\infty}$) where the pair susceptibility diverges. By calculating the pair fluctuation propagator $L(\bs q, \omega)$ and partition-function, we determine the spectral density in the normal state. We find an interaction-driven, power-law $\frac{1}{\sqrt{\omega}}$ type, van-Hove singularity (vHS) at low energies that signals NFL physics. Hence pair fluctuations combined with LS physics describe a NFL-SC transition at $T=0$. \textcolor{black}{As a key feature we find that, in the strong coupling limit ($\beta u \gg 1$), the leading order fluctuation free energy terms in the normal state of this NFL-SC transition takes a form similar to the equivalent $O(1)$ free energy terms of the Sachdev-Ye-Kitaev (SYK) model. This free energy contribution is given by $-\beta F = \beta u_{c\infty} - \gamma~\text{ln}\left(\beta u_{c \infty}\right)$ where $\gamma = \frac{1}{2}$.} Here $u$ is the interaction parameter and is equal to square-root of the residue of the self-energy pole. Moreover, we do not require random couplings for our conclusions to hold. In the presence of weak impurity scattering, the low-energy spectral density is unaffected in the strong coupling limit and gives rise to an interaction-driven enhancement of superconductivity.  \textcolor{black}{Our results point toward order-parameter fluctuations acting as conduits between Mott physics and strongly coupled 
toy-models exhibiting gravity duals~\cite{Stanford2016, Pochinski2017}.} \par
\begin{figure}
\includegraphics[width=3.5in,height=1.4in]{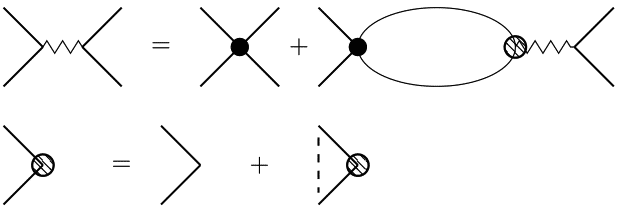}
\caption{Bethe-Salpeter equations for the fluctuation propagator (denoted by zig-zag lines) in the particle-particle channel. The shaded (black solid) disk denotes vertex corrections due to impurities (bare interaction). The dashed (solid) lines denote impurity scatterers (electron Green function).} \label{Feynman}
\end{figure}
\section{Model}
LSs have been obtained in numerous models in many-body literature, both at a phenomenological level~\cite{Zhang2006} as well as microscopic Hubbard-type~\cite{Tsvelik2002, Tsvelik2006, Kotliar2006,Rosch2007, Imada2009, Ohta2011, Hong-Phillips2012, Kane2013, Sachdev2018} and holographic~\cite{Phillips-Leigh2011-1, Phillips-Leigh2011-2} models. Other models study emergent gauge fields in a FL that nevertheless violate the LSR ~\cite{Sachdev2016, Vojta2003, Wen2012, Sachdev2012}.  The simplest Green function that vanishes along contours in the Brillouin zone has a simple pole in the self-energy and is given by $G(\bs p, \epsilon_n)^{-1} = i \epsilon_n - \xi(\bs p) - \Sigma(\bs p, \epsilon_n) $, where $\xi(\bs p) = \epsilon(\bs p) - \mu$ is the bare dispersion with chemical potential $\mu$ and $\epsilon_n$ is the fermionic Matsubara frequency. We choose a self-energy ansatz motivated by the well-studied Yang-Rice-Zhang model (YRZ)~\cite{Zhang2006} with a pole structure given by 
\beq
\Sigma(\bs p, i \epsilon_n) =  \frac{u^2}{i \epsilon_n + \xi(\bs p)}.
\eeq
\textcolor{black}{The constant Hartree-Fock potential is dropped as it results in a trivial renormalization of the bands}. As evident from the choice of $\Sigma$, the LS and the bare electron FS occur for the same momenta set by $\mu$ at zero energy. This need not be the case in more generic systems where the self-energy can acquire multiple poles each with distinct residues. \textcolor{black}{The square of the interaction, $u^2$, is a quantity determined by microscopic parameters of the Hamiltonian.} In the presence of impurities, a finite life-time $\tau$ is introduced in the Green function. \textcolor{black}{While the YRZ Green function was motivated by a doped spin-liquid, it is nevertheless a popular example of a minimal (phenomenological) model that has a  diverging self-energy; hence it forms a workable model of a Luttinger surface. The Bethe-Salpeter equation appearing in Fig.~\ref{Feynman} is an equation for vertex correction and is used to define the bosonic fluctuation propagator in the theory of fluctuation superconductivity~\cite{LarkinVarlamov2005}. It assumes a knowledge of the full electronic Green function as an input, either obtained phenomenologically, or from a microscopic Hamiltonian. While in the conventional theory of fluctuations, the pair-bubble is approximated by use of the Green function of non-interacting electrons, we use the aforementioned YRZ proposal  as our phenomenological input (see also end of Discussion section and the note added after). } \par 
\textit{Strong coupling ($\beta u\gg 1$) in clean limit ($\tau\rightarrow \infty$):} The fluctuation propagator can be evaluated from Bethe-Salpeter-type equations in the particle-particle channel (see Fig.~\ref{Feynman}) for momentum $\bs q$ and frequency $\Omega$ as
\beq
L^{-1}(\bs q, \Omega) = -g^{-1} + \Pi(\bs q, \Omega)
\eeq
 where $-g$ is a constant bare (attractive) interaction vertex and  $\Pi(\bs q, \Omega)$ is the pair susceptibility. The latter is defined in $d$-dimensions as $  \Pi(\bs q, \Omega_k) = \frac{1}{\beta (2\pi)^d} \sum_{\epsilon_n} \int d^d\bs p~G(\bs p + \bs q, \epsilon_{n+k})G(-\bs p, -\epsilon_{n})$, \textcolor{black}{where $\epsilon_n \equiv (2 n + 1) \pi T$ and $\Omega_k \equiv 2 k \pi T$ are the fermionic and bosonic Matsubara frequencies}. Substituting $G(\bs p, \epsilon_n)$ into $ \Pi(\bs q, \Omega)$ and taking the limit of $|\bs q| \equiv q \ll p_f$, the Fermi momentum, we obtain for quadratic  bands in $d=2$ (see Appendix A)
 \beq \nonumber
 \Pi(\bs q, \Omega_k) &=& \Bigg\langle \frac{m}{2 \beta} \sum_n \Big[ \frac{(\epsilon_1 + \epsilon_1')(\epsilon_1'+ \epsilon_2 + i r c_{\phi})}{\epsilon_1' (\epsilon_1^2 - \epsilon_2^2 - r^2 c_{\phi}^2 + 2 i c_{\phi} r \epsilon_1')}  \nonumber \\
&&  + \text{c.c}~(1\leftrightarrow 2)\Big]\Bigg\rangle, \label{AngularAverage}
 \eeq
where we make the replacements $\epsilon_1\equiv \epsilon_{1n}\rightarrow \epsilon_{n+k}$ and $\epsilon_2 \equiv \epsilon_{2n}\rightarrow -\epsilon_n$, and introduce primed notation $\epsilon_{jn}' \equiv\sqrt{\epsilon_{jn}^2 + u^2} $. The angular brackets $\langle ...\rangle$ denote angular average, and $m$ and $\phi$ are the bare electron mass and azimuthal angle respectively. \textcolor{black}{To recover well-known expressions of the pair-susceptibility for a FL, one only needs to take the limit of $u\rightarrow 0$ (see Appendix B). We now introduce the ratio $r= \frac{p_f q}{m}\ll u, \epsilon_f$, where $\epsilon_f$ is the Fermi energy.  Performing an expansion in the parameter $r/u$ (strong interactions) and taking the static limit}, the inverse fluctuation propagator is
\beq \label{FluctuationPropagator}
L^{-1}(\bs q, \Omega\rightarrow 0) &\simeq& -g^{-1} + \Pi^{(0)}(0,0) + \Pi^{(2)}(\bs q,0), \\ \nonumber
\eeq
where $\Pi^{(0)}(0,0) = \frac{m}{4}\left(2 S_1 - u^2 S_3\right)$, $\Pi^{(2)}(\bs q,0) = -\frac{m r^2}{32}\left(2 S_3 - u^2 S_5\right)$ and  $S_{\nu} = \frac{1}{\beta} \sum_{\epsilon_n} (\epsilon_n^2 + u^2 )^{-\nu/2} $. These sums can be evaluated exactly for odd $\nu$ and we obtain for $\Lambda \gg u \gg T$ (see Appendix C) 
\begin{widetext}
\begin{align}
S_{\nu} =\begin{cases}
    \text{$\frac{1}{\pi} \ln\frac{\Lambda}{u} - \frac{2}{\pi} K(0,\kappa)$}~~~~~~~~~~~~~~~~~~~~~~~~~~~~~~~~~~~~~~~~~~~~~~~~~~~~~~~\text{for $\nu=1$ }\\
       \text{$\frac{e^{- i \pi \nu} \Gamma\left(1- \frac{\nu}{2}\right) \sin \pi \left( 1- \frac{\nu}{2} \right)}{u^{\nu-1}2 \pi^{3/2}}\left[ 2^{\frac{5-\nu}{2}} \kappa^{\frac{\nu -1}{2}} K\left( \frac{\nu-1}{2}, \kappa\right) - \Gamma\left(\frac{\nu -1}{2}\right) \right]$}~~~~~~\text{for $\nu = 3, 5, ..$.} 
     \end{cases} \label{Minimum}
\end{align}
\end{widetext}
Here $\Lambda$ is the ultraviolet cut-off of the divergent Matsubara sum for $\nu =1$ (plays the role of the Debye frequency $\omega_D$ in the conventional BCS theory), $\kappa \equiv \beta u$, $\Gamma(x)$ is the gamma function and $K(x,y)$ is the modified Bessel function of the second kind. Substituting for $S_{\nu}$ into Eq.~\ref{FluctuationPropagator} for the inverse fluctuation propagator, and expanding the resulting expression in powers of $e^{-\kappa}$ and its polynomial products, we obtain the final expression for $L^{-1}(\bs q, 0)$ in the clean limit
\beq \nonumber
L^{-1}(\bs q, \Omega\rightarrow 0) &\simeq& - g^{-1} + N_0 \left[\ln\frac{\Lambda}{u} + \sqrt{\frac{\pi \kappa}{2}} e^{-\kappa}\right] \\
&& - \frac{N_0 r^2}{12 u^2} \left[ 1 + \sqrt{\frac{\pi \kappa^3}{8}} e^{-\kappa} \right].
\label{NonCriticalPropagator}
\eeq
Here $N_0$ is the density of states at the Fermi level in two dimensions. Note that the above expression cannot be adiabatically connected to the FL result~\cite{LarkinVarlamov2005} any longer as it is valid only in the strong coupling limit. There are several conclusions that can be drawn from the structure of the fluctuation propagator above. First, a divergence of the zero frequency, long-wavelength limit of the propagator signals a superconducting instability. At $\beta = \infty$ and constant $\Lambda$, this condition is achieved at the quantum critical point $u=u_{c\infty} = \Lambda~e^{-\frac{1}{N_0 g}}$, a form analogous to the thermal BCS-type transition. Hence, interactions can destroy superconductivity even at zero temperature if $u>u_{c\infty}$. \textcolor{black}{This must be contrasted with a recent result~\cite{Schmalian2019} where SYK-type random electron-phonon couplings lead to pairing instead of the low temperature quantum critical state. The quantum critical point in the current model can indeed be avoided yielding superconducting pairing provided the surface of zeros is partially or fully destroyed.} 
Second, the conformal structure of the theory is highlighted by setting $u=u_{c\infty}$ where the static, long-wavelength propagator \textcolor{black}{at low but non-zero temperatures} takes a familiar form
\beq \label{CriticalPropagator}
L^{-1}(\bs q\rightarrow0, \Omega = 0)_{u=u_{c\infty}} = N_0 \sqrt{\frac{\pi u_{c\infty}}{2 T}} e^{\frac{-u_{c\infty}}{T}}
\eeq
 From this expression, it is illuminating to evaluate the fluctuation contribution to the free energy to zeroth order in $\bs q$  above the critical point. Following the procedures described in~\cite{LarkinVarlamov2005, Vinokur2005} for the case of a single band model and using Eq.~\ref{CriticalPropagator}, we obtain the \textcolor{black}{leading order} pair fluctuation free energy
 \beq\label{FreeEnergy}
 -\beta F = \beta u_{c\infty} - \gamma~\ln(\beta u_{c\infty}).
 \eeq
\begin{figure}[h!]
\includegraphics[width=1.6in,height=1.6in]{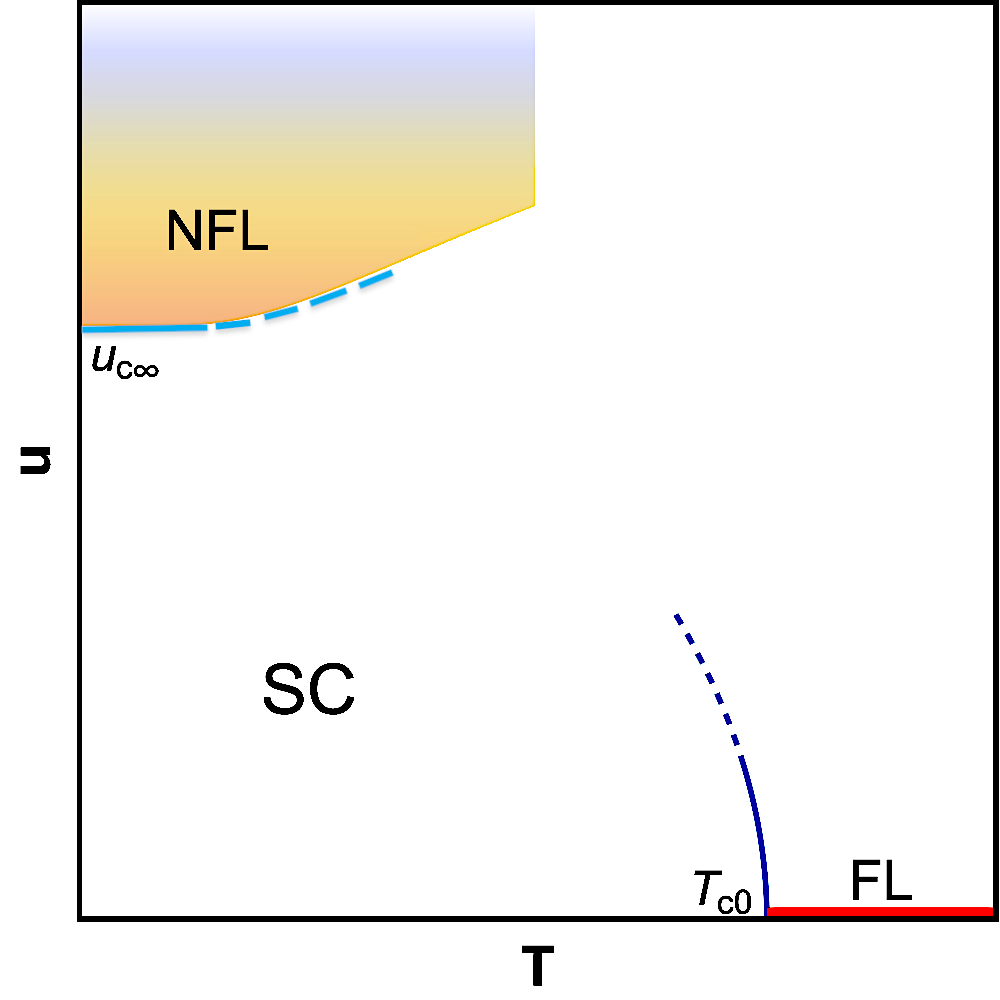} \hfill
\includegraphics[width=1.75in,height=1.6in]{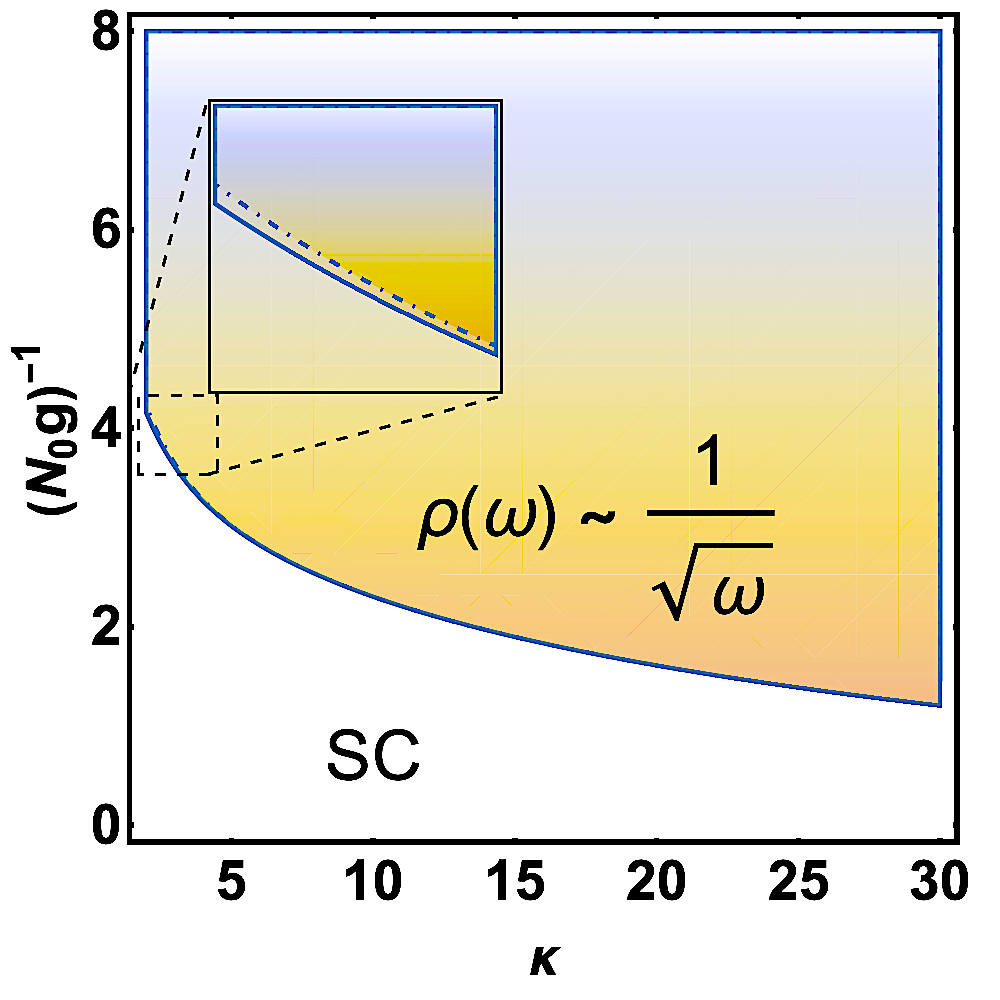}
\caption{(Left) Schematic plot of the $u$-$T$ phase diagram in the clean limit. The red solid line denotes a Fermi liquid ($u=0$) while the light (dark) blue contours define the phase boundary in the strong coupling $\beta u \equiv \kappa \gg 1$ (weak coupling $\beta u \ll 1$) limit. The strong coupling normal state is a NFL with a power-law divergence of the spectral density $\rho(\omega)$. We have defined $T_{c0} \equiv T_c(u=0)$ and $u_{c\infty} \equiv u_c(\beta\rightarrow \infty)$. (Right) Strong coupling, weak impurity scattering ($T\tau\equiv \theta\gg1$) limit of the $\kappa$-$(N_0 g)^{-1}$ phase diagram. On the solid (dotted-dashed) curve, the pair fluctuations diverge in the absence (presence) of impurity scattering.  Inset shows the weak enhancement of SC phase due to impurities.} \label{PD}
\end{figure}
 where $\gamma=\frac{1}{2}$. This result must be compared with other quantum critical models having gravity duals such as the Sachdev-Ye-Kitaev (SYK) model and its variants~\cite{Kitaev2015,Ye1993, Sachdev2010, Stanford2016, Witten2016} where \textcolor{black}{the leading order fluctuation contribution to the free energy about the saddle point} acquires a form similar to Eq.~\ref{FreeEnergy}, but with $\gamma=\frac{3}{2}$~\cite{Stanford2016, Pochinski2017}. \textcolor{black}{ It is useful to note that the calculated free energy terms should be compared to the $O(N^0)$ terms in the SYK model (where $N$ is the number of fermionic flavors) as these are the only contributions that arise from leading order fluctuations about the respective saddle point solutions.} \textcolor{black}{Moreover, the fluctuation free energy terms match the SYK results only for the $\bs q\rightarrow 0$ mode where the spatial dimensionality is smeared out, consistent with an effective zero dimensional model. Hence, and as should be anticipated, it is only in the long-wavelength limit that the proposed mapping of the fluctuation free energy to the $O(N^0)$ SYK model holds.}  Finally, one can evaluate the spectral density $\rho(\omega)$ by taking the inverse Laplace transform of the partition function and the resulting integral can be solved by the saddle point method~\cite{Stanford2016}. While $\rho(\omega)$ is a constant independent of $\omega$ at low energies in the SYK-type models~\cite{Stanford2016}, our model yields a vHS $\rho(\omega)\sim \frac{1}{\sqrt{\omega}}$ at low energy leading to NFL transport~\cite{Wenbo-Sachdev2017}. This contrast is entirely due to the difference in the coefficient $\gamma$ of the log term in Eq.~\ref{FreeEnergy}. The conclusions drawn above are summarized in Fig.~\ref{PD}. The $u$-$T$ phase diagram in Fig.~\ref{PD} (left panel) plots the strong coupling phase boundary (solid light blue line) separating the SC and NFL phases for a constant $\Lambda$.  The dashed lines are extrapolations of the phase boundary where approximations made above fail.  Fig.~\ref{PD} (right panel, solid curve) plots the $\kappa$-$(N_0 g)^{-1}$ phase diagram and shows the same phase boundary for close to zero temperatures and constant $\beta \Lambda$. The intensity of fluctuations is indicated by the color scale and is largest in magnitude right above the phase boundary. \par
\textit{Weak coupling ($\kappa = \beta u\ll 1$) in clean limit ($\tau\rightarrow \infty$):} That the $T=0$ pair instability is only a feature  at strong coupling can be confirmed by calculating $L^{-1}(\bs q, 0)$ in the opposite (weak coupling) limit $\beta u\ll 1$. We begin with Eq.~\ref{FluctuationPropagator} and expand $ \Pi^{(0)}(0,0)$ and $ \Pi^{(2)}(\bs q,0)$ to quadratic power in $\kappa =\beta u$ to obtain
\beq 
\Pi^{(0)}(0, 0)&\simeq& \frac{m}{4\beta} \sum_{\epsilon_n} \left[ \frac{2}{|\epsilon_n|} - \frac{2 u^2}{|\epsilon_n|^3} \right] \\ 
 \Pi^{(2)}(\bs q,0) &\simeq& \frac{-m r^2}{32 \beta}\sum_{\epsilon_n} \left[ \frac{2}{|\epsilon_n|^3} - \frac{4 u^2}{|\epsilon_n|^5}\right].
\eeq 
The sums above can be performed and substituted back into the static limit of the propagator (see Appendix D) and we find,
\beq\nonumber
L^{-1}(\bs q, \Omega\rightarrow0) &=& -\frac{1}{g} + N_0 \left[  \ln~\frac{\Lambda}{2\pi T} - \psi\left(\frac{1}{2}\right) - \frac{u^2 C_2}{8\pi^2T^2}\right] \\
&&-\frac{N_0 r^2}{128 \pi^2 T^2} \left[ 2 C_2- \frac{u^2 C_4}{12 \pi^2 T^2} \right],
\eeq
where $C_2 = |\psi''\left(  \frac{1}{2}\right)|$, $C_4 = |\psi^{(4)}\left(  \frac{1}{2}\right)|$ are numerical constants equal to the second and fourth derivatives of the digamma function $\psi(x)$ respectively. Setting $\bs q \rightarrow 0$, this form of  the fluctuation propagator resembles its thermal BCS counterpart plus the correction term proportional to $(\beta u)^2$. It is hence clear that there is no sensible way to obtain a zero temperature transition into the superconducting state (since $\beta u\ll 1$). Moreover, as the correction term is negative, its effect on BCS result is to reduce the thermal transition temperature $T_c$ for a given interaction strength $g$ and energy cut-off $\Lambda$. This is shown in Fig.~\ref{PD} (left panel) where we have defined $T_{c0} \equiv T_c(u=0)$ and the dashed lines are extrapolations of the phase boundary where approximations made above fail.
 \par
\textit{Strong coupling ($\beta u \gg 1$) and dilute impurities ($\theta \equiv T\tau \gg 1$):} The fluctuation propagator in the presence of impurities is shown in Fig.~\ref{Feynman} -- the solid lines are now impurity Green functions that acquire zeros and the shaded disk denotes vertex corrections due to impurities. The pair susceptibility bubble then becomes~\cite{LarkinVarlamov2005} ($d=2$) 
\beq\label{ImpurityPairSusceptibility}
\Pi(\bs q, \Omega_k) = \frac{1}{\beta}\sum_{\epsilon_n} \frac{P(\bs q, \tilde{\epsilon}_{n+k}, -\tilde{\epsilon}_{n})}{1 - \frac{P(\bs q, \tilde{\epsilon}_{n+k}, -\tilde{\epsilon}_{n})}{2\pi N_0 \tau}},
\eeq
where $P(\bs q, \tilde{\epsilon}_{1}, \tilde{\epsilon}_{2}) =  \frac{1}{(2\pi)^2} \int d^2\bs p~G(\bs p+ \bs q, \tilde{\epsilon}_1) G(-\bs p, \tilde{\epsilon}_2)$, $\tilde{\epsilon}_n = \epsilon_n + \frac{\text{sgn}(\epsilon_n)}{2\tau}$, and sgn$(x)$ is the sign function. For $\Omega_k =0$, one can perform an expansion in $r$ similar to the clean case and write 
\beq \nonumber
P(\bs q, \tilde{\epsilon}_n, -\tilde{\epsilon}_n) &\simeq& P^{(0)}(\bs q =0, \tilde{\epsilon}_n, -\tilde{\epsilon}_n) + P^{(2)}(\bs q, \tilde{\epsilon}_n, -\tilde{\epsilon}_n), \\
 P^{(0)}(0,0) &=& \frac{\beta m}{4}\left(2 \tilde{S}_1 - u^2 \tilde{S}_3\right), \\
  P^{(2)}(\bs q,0) &=& -\frac{\beta m r^2}{32}\left(2 \tilde{S}_3 - u^2 \tilde{S}_5\right)
\eeq
 and  $\tilde{S}_{\nu} = \frac{1}{\beta} \sum_{\epsilon_n} (\tilde{\epsilon}_n^2 + u^2 )^{-\nu/2} $. In the limit $\beta u \gg 1$ and $\theta \equiv T\tau \gg 1$, the denominator in Eq.~\ref{ImpurityPairSusceptibility} can be approximated by unity. This is equivalent to ignoring vertex corrections due to impurity scattering and hence $\Pi(\bs q, \Omega_k=0) \simeq \frac{1}{\beta} \sum_n P(\bs q, \tilde{\epsilon}_n, -\tilde{\epsilon}_n)$ (Appendix G gives additional numerical justification for this approximation). The  Matsubara sums can be performed exactly for $u>\tau^{-1}$ (Appendix F) and the final expression for the fluctuation propagator is only slightly modified from the clean limit and given by
\beq \nonumber
L^{-1}(\bs q, \Omega\rightarrow 0) &\simeq& - g^{-1} + N_0 \left[\ln\frac{\Lambda}{u} + \sqrt{\frac{\pi \kappa}{2}} e^{-\kappa + \frac{1}{2\theta}} \right] \\
&& - \frac{N_0 r^2}{12 u^2} \left[ 1 + \sqrt{\frac{\pi \kappa^3}{8}} e^{-\kappa + \frac{1}{2 \theta}} \right].
\eeq
Hence its conformal structure at the quantum critical point, $L^{-1}(\bs q\rightarrow0, \Omega = 0)_{u=u_{c\infty}} = N_0 \sqrt{\frac{\pi u_{c\infty}}{2 T}} e^{\frac{-u_{c\infty}}{T} + \frac{1}{2\theta}}$, as well as the free energy contribution and vHS in $\rho(\omega)$ are left essentially unchanged. On the other hand, as shown in Fig.~\ref{PD} (right panel), there is a weak \textit{enhancement} of the superconducting phase in the strong coupling phase diagram.\par
\textit{Weak coupling ($\beta u \ll 1$) and dilute impurities ($T\tau\gg1$):} The final case we consider is the weak coupling limit in the presence of dilute impurities. In this limit, vertex corrections become more important than in the strong coupling case (Appendix G) and the static long-wavelength limit of the pair susceptibility is
\beq
\Pi(\bs q \rightarrow 0, \Omega =0) &=& \frac{1}{\beta}\sum_{n} \frac{2\pi N_0 \tau \tilde{A}(\tilde{\epsilon}_n, u)}{2\pi N_0 \tau - \tilde{A}(\tilde{\epsilon}_n, u)}\\
\tilde{A}(\tilde{\epsilon}_n, u) &=& \frac{2 \pi N_0}{4}\left[ \frac{2}{\tilde{\epsilon}_n'} - \frac{u^2}{\tilde{\epsilon}_n'^3}\right].
\eeq  
Like in the case of the clean limit, we can perform an expansion in $\beta u$  and the Matsubara summations have been performed in Appendix E. The final result for the fluctuation propagator in this limit takes the form
\begin{widetext}
\begin{align}
L^{-1}(\bs q \rightarrow 0, \Omega =0)  = -g^{-1} + N_0\left[ \ln~\left(\frac{\Lambda}{4\pi T}\right) - \psi\left(\frac{1}{2}\right) + 4 u^2 \tau^2 \left[  \psi\left(\frac{1}{2} + \frac{1}{4\pi \theta}\right)  -\frac{1}{4\pi\theta} \psi'(\frac{1}{2}) - \psi\left(\frac{1}{2}\right)\right]\right].
\end{align}
\end{widetext}
This expression for the propagator looks similar to that obtained in the limit of low $\bs q$ and $\tau<\infty$, but with $u^2$ replacing the energy scale arising from the squared momentum factor~\cite{LarkinVarlamov2005}. In this limit, as anticipated from the clean case, the conformal structure of the propagator is lost and there is only a thermal transition into the superconducting state. \par
\section{Discussion}
\textcolor{black} {As alluded to in the main text, the near-conformal structure of the fluctuation propagator and the associated free energy  (Eqs.~\ref{CriticalPropagator} and \ref{FreeEnergy}) obtained from the YRZ Green function is reminiscent of the SYK model discussed in Ref.~\cite{Stanford2016}. The difference in the coefficient $\gamma$ of the logarithmic term results in a power-law divergence of the spectral density as opposed to a constant as in the SYK model. }  
 Moreover, the physics in both cases is controlled by a single parameter in the strong coupling-low temperature limit. In the SYK case, it is the parameter $\beta J \rightarrow \infty$ whereas in the YRZ case its is $\beta u\rightarrow \infty$. \textcolor{black}{In both scenarios, the respective dualities -- SYK to 2-dimensional gravity and YRZ fluctuations to SYK (to leading order in fluctuations)-- occur only in the strong coupling limit.} Furthermore, although the SYK Green function is \textit{local}, given by $G(i\omega_n) = -i \omega_n - \Sigma(i\omega_n)$, as opposed to the momentum dependent YRZ case, the respective free energy terms coincide only for the $\bs q\rightarrow 0$ YRZ mode where the spatial dimensionality is smeared out, reducing the model to an effectively zero dimensional system like the SYK.  \textcolor{black}{The key difference, however, is the absence of any disorder (a result consistent with the conclusion of Ref.~\cite{Witten2016})} needed in our calculations; instead, we require a momentum-dependent (non-local) self-energy to obtain the same physical content. In addition, a large-$N$ parameter, typically used in SYK-type models, is absent. We also emphasize that the equivalence between Eqs.~\ref{CriticalPropagator}, \ref{FreeEnergy} and the corresponding quantities in gravity-type models holds without additional vertex corrections from Coulomb interactions. Therefore, these terms are expected to be unimportant for establishing this equivalence (\textcolor{black}{see also note added at the end}). Furthermore, the weak enhancement of the superconducting phase induced by the interplay of electron correlations and dilute impurities is also consistent with previous studies~\cite{Seki1995, Larkin2002, Setty2019}. Looking ahead, it would be of considerable interest to examine the consequences of fluctuation-driven vHS on properties such as the entanglement entropy and energy-level spacing near the quantum critical point for a model with a LS. \textcolor{black}{Additionally, the difference in the factor $\gamma$ in the fluctuation free energy appearing in Eq.~\ref{FreeEnergy} suggests a generalization of the Schwarzian action describing in the AdS$_2\times$S$_2$ geometry~\cite{Stanford2016}. This could provide crucial insight into this distinction between fluctuations on a LS and SYK. Finally, while no explicit reference to a microscopic model has been made in this work, the conclusions in this paper are applicable to exactly solvable Hamiltonians hosting LSs such as those in Ref.~\cite{Kohmoto1992}. This is the subject of ongoing and future work. }\par
 \textcolor{black}{\textit{Note Added:} After the completion of this work, a follow-up preprint by Phillips and coworkers~\cite{Phillips2019} pointed out that an LS of the form similar to the one obtained through the YRZ Green function could also be realized via an exactly solvable microscopic model with long-range interactions by Hatsugai and Kohmoto~\cite{Kohmoto1992}. The results appearing in the current paper are applicable to this model in the presence of an additional superconducting interaction term~\cite{Phillips2019} and will be addressed in a forthcoming manuscript.  } \\ \newline
\textit{Acknowledgements:} We thank P. W. Phillips and G. La Nave for illuminating discussions and B. Padhi for a critical reading of the manuscript. This work is supported by the DOE grant number DE-FG02-05ER46236.

 \bibliographystyle{apsrev4-1}
\bibliography{LuttingerSurface.bib}
\newpage
\onecolumngrid
\section{Appendix A}
In this Appendix, we derive expression for the pair susceptibility $\Pi(\bs q, \Omega_k)$ appearing in Eq.~\ref{AngularAverage} of the main text. For the clean limit in $d=2$ we begin with the definition ($\epsilon_1 \equiv \epsilon_{1n}, \epsilon_2 \equiv \epsilon_{2n} $) 
\beq
 I(\bs q, \epsilon_1, \epsilon_2) = \int d^2\bs p~G(\bs p + \bs q, \epsilon_1)G(-\bs p, \epsilon_2)
\eeq
where 
\beq
G(\bs p, \epsilon_n) &=& \frac{1}{i\epsilon_n - \frac{p^2}{2m} + \mu -\Sigma(\bs p, \epsilon_n) },\\
\Sigma(\bs p, \epsilon_n) &=& V + \frac{u^2}{i\epsilon_n + \xi(\bs p)},
\eeq
$\xi(\bs p) = \epsilon(\bs q) - \mu$, and $V$ is a constant potential. $u^2$ is the residue of the self-energy pole and its square-root plays the role of an interaction strength to which other quantities can be compared. Substituting $\Sigma(\bs p, \epsilon_n) $ back into $I(\bs q, \epsilon_1, \epsilon_2)$ and taking the limits $|\bs q| \equiv q\ll p_f$, $\omega \equiv \epsilon_1 - \epsilon_2 \sim \frac{p_f q}{m} \ll \frac{p_f^2}{2m}$ we get
\beq
 I(\bs q, \epsilon_1, \epsilon_2) = m \int_{-\infty}^{\infty} dx \int_0^{2\pi} d\phi \frac{\left(i\epsilon_1 + x + r cos\phi \right)\left(i\epsilon_2+x \right)}{\left((i\epsilon_2)^2 - x^2 - u^2 \right)\left( (i \epsilon_1)^2 - (x + r cos\phi)^2- u^2 \right)}.
 \eeq
To obtain the above, we have made the replacements $x = \frac{p^2}{2 m} - \mu$, $\int d^2\bs p = m \int d\left(\frac{p^2}{2 m} \right) d\phi$, and absorbed $V$ into the definition of the chemical potential which is set to be large. The poles of the integrand in $ I(\bs q, \epsilon_1, \epsilon_2)$ are located at $x = \pm i \sqrt{\epsilon_2^2 + u^2} \equiv \pm i \epsilon_2'$ and $\pm i \sqrt{\epsilon_1^2 + u^2} - r cos\phi \equiv \pm i \epsilon_1' - r cos\phi$. Along side these definitions, we can write the pair susceptibility as 
\beq
\Pi(\bs q, \Omega_k)  &=& \frac{1}{\beta (2\pi)^2} \sum_{\epsilon_n}  I(\bs q, \epsilon_1, \epsilon_2) \\
&=& \left \langle \frac{m}{2\pi \beta} \sum_{\epsilon_n} \int_{-\infty}^{\infty} dx \left[ \frac{\left(i\epsilon_1 + x + r cos\phi \right)\left(i\epsilon_2+x \right)}{\left((i\epsilon_2)^2 - x^2 - u^2 \right)\left( (i \epsilon_1)^2 - (x + r cos\phi)^2- u^2 \right)} \right]\right \rangle,
\eeq
with the replacements $\epsilon_1 \rightarrow \epsilon_{n+k}$ and $\epsilon_2 \rightarrow -\epsilon_n$ and the angular average $\langle ... \rangle \equiv \frac{1}{2\pi} \int_0^{2\pi} d\phi~$. The integral over the variable $x$ can be performed exactly by the method of residues. Using poles of the $x$ integrand and summing over residues in the upper-half plane, we obtain the pair susceptibility as
\beq
\Pi(\bs q, \Omega_k)&=&  \left \langle \frac{m}{2\pi \beta} \sum_{\epsilon_n} \left[\frac{\pi\left(\epsilon_1 + \epsilon_1'\right) \left( \epsilon_2 + \epsilon_1' + i r cos\phi\right)}{\epsilon_1'(\epsilon_1^2 - \epsilon_2^2 - r^2 cos^2\phi + 2 i r \epsilon_1' cos\phi)} + \frac{\pi\left(\epsilon_2 + \epsilon_2'\right) \left( \epsilon_1 + \epsilon_2' - i r cos\phi\right)}{\epsilon_2'(\epsilon_2^2 - \epsilon_1^2 - r^2 cos^2\phi - 2 i r \epsilon_2' cos\phi)} \right]\right \rangle, \\
&=& \left \langle \frac{m}{2\pi \beta} \sum_{\epsilon_n} \left[\frac{\pi\left(\epsilon_1 + \epsilon_1'\right) \left( \epsilon_2 + \epsilon_1' + i r cos\phi\right)}{\epsilon_1'(\epsilon_1^2 - \epsilon_2^2 - r^2 cos^2\phi + 2 i r \epsilon_1' cos\phi)} + c.c (1\leftrightarrow 2) \right]\right \rangle.
\eeq
This is the expression that appears in Eq.~\ref{AngularAverage} of the main text. 
\section{Appendix B}
In this Appendix, we show that Eq.~\ref{AngularAverage} of the main text indeed reduces to the correct FL result in the limit $u\rightarrow0$. We begin with the expression for $\Pi(\bs q, \Omega_k)$ (we make the replacements $\epsilon_1\equiv \epsilon_{1n}\rightarrow \epsilon_{n+k}$ and $\epsilon_2 \equiv \epsilon_{2n}\rightarrow -\epsilon_n$ to recover the $\Omega_k$ dependence)
\beq \nonumber
\Pi(\bs q, \Omega_k)&=&  \left \langle \frac{m}{2\pi \beta} \sum_{\epsilon_n} \left[\frac{\pi\left(\epsilon_1 + \epsilon_1'\right) \left( \epsilon_2 + \epsilon_1' + i r cos\phi\right)}{\epsilon_1'(\epsilon_1^2 - \epsilon_2^2 - r^2 cos^2\phi + 2 i r \epsilon_1' cos\phi)} + \frac{\pi\left(\epsilon_2 + \epsilon_2'\right) \left( \epsilon_1 + \epsilon_2' - i r cos\phi\right)}{\epsilon_2'(\epsilon_2^2 - \epsilon_1^2 - r^2 cos^2\phi - 2 i r \epsilon_2' cos\phi)} \right]\right \rangle.
\eeq
Noting that all square-roots appearing above are positive, we have $\epsilon_i' \rightarrow |\epsilon_i|$ as $u\rightarrow0$. Hence, in this limit we have
\beq \nonumber
\Pi(\bs q, \Omega_k)_{u\rightarrow 0} =  \left \langle \frac{m}{2\pi \beta} \sum_{\epsilon_n} \left[\frac{\pi\left(\epsilon_1 + |\epsilon_1|\right) \left( \epsilon_2 + |\epsilon_1| + i r cos\phi\right)}{|\epsilon_1|(\epsilon_1^2 - \epsilon_2^2 - r^2 cos^2\phi + 2 i r |\epsilon_1| cos\phi)} + \frac{\pi\left(\epsilon_2 + |\epsilon_2|\right) \left( \epsilon_1 + |\epsilon_2| - i r cos\phi\right)}{|\epsilon_2|(\epsilon_2^2 - \epsilon_1^2 - r^2 cos^2\phi - 2 i r |\epsilon_2| cos\phi)} \right]\right \rangle.
\eeq
\textit{Case 1}, $\epsilon_1<0; \epsilon_2<0$: The numerators of both the terms vanish since $\epsilon_i + |\epsilon_i| = -|\epsilon_i| + |\epsilon_i|  =0$, hence $\Pi(\bs q, \Omega_k)_{u\rightarrow 0} = 0$ for this case. \newline
\textit{Case 2}, $\epsilon_1>0; \epsilon_2>0$: In this case, both the numerators are non-zero but the two terms cancel, i.e., 
\beq
\Pi(\bs q, \Omega_k)_{u\rightarrow 0} = 2 \pi \left[ \frac{|\epsilon_2| + |\epsilon_1| + i r cos\phi}{(|\epsilon_1| + i r cos\phi)^2 - \epsilon_2^2} -  \frac{|\epsilon_1| + |\epsilon_2| - i r cos\phi}{ \epsilon_1^2 - (|\epsilon_2| - i r cos\phi)^2 } \right] = 0.
\eeq
\textit{Case 3}, $\epsilon_1<0; \epsilon_2>0$: Here, the first term equals zero but the second remains non-zero and we have 
\beq
\Pi(\bs q, \Omega_k)_{u\rightarrow 0} = 2 \pi \frac{|\epsilon_1| - |\epsilon_2| + i r cos\phi}{\epsilon_1^2 - (|\epsilon_2| - i r cos\phi)^2} = \frac{2\pi}{|\epsilon_1| + |\epsilon_2| - i r cos\phi}.
\eeq
\textit{Case 4}, $\epsilon_1>0; \epsilon_2<0$: Similar to the case above, we have a non-zero contribution from the first term to give
\beq
\Pi(\bs q, \Omega_k)_{u\rightarrow 0} = 2 \pi \frac{|\epsilon_1| - |\epsilon_2| + i r cos\phi}{-\epsilon_2^2 + (|\epsilon_1| + i r cos\phi)^2} = \frac{2\pi}{|\epsilon_1| + |\epsilon_2| + i r cos\phi}.
\eeq
We can combine all the cases above to write 
\beq
\Pi(\bs q, \Omega_k)_{u\rightarrow 0} = \frac{2\pi \Theta(-\epsilon_1 \epsilon_2)}{|\epsilon_1 - \epsilon_2| + i sgn(\epsilon_1 - \epsilon_2) r cos\phi},
\eeq
which is the same as the expression derived for the FL case~\cite{LarkinVarlamov2005}.
\section{Appendix C}
\begin{figure}[h!]
\includegraphics[width=3in,height=3in]{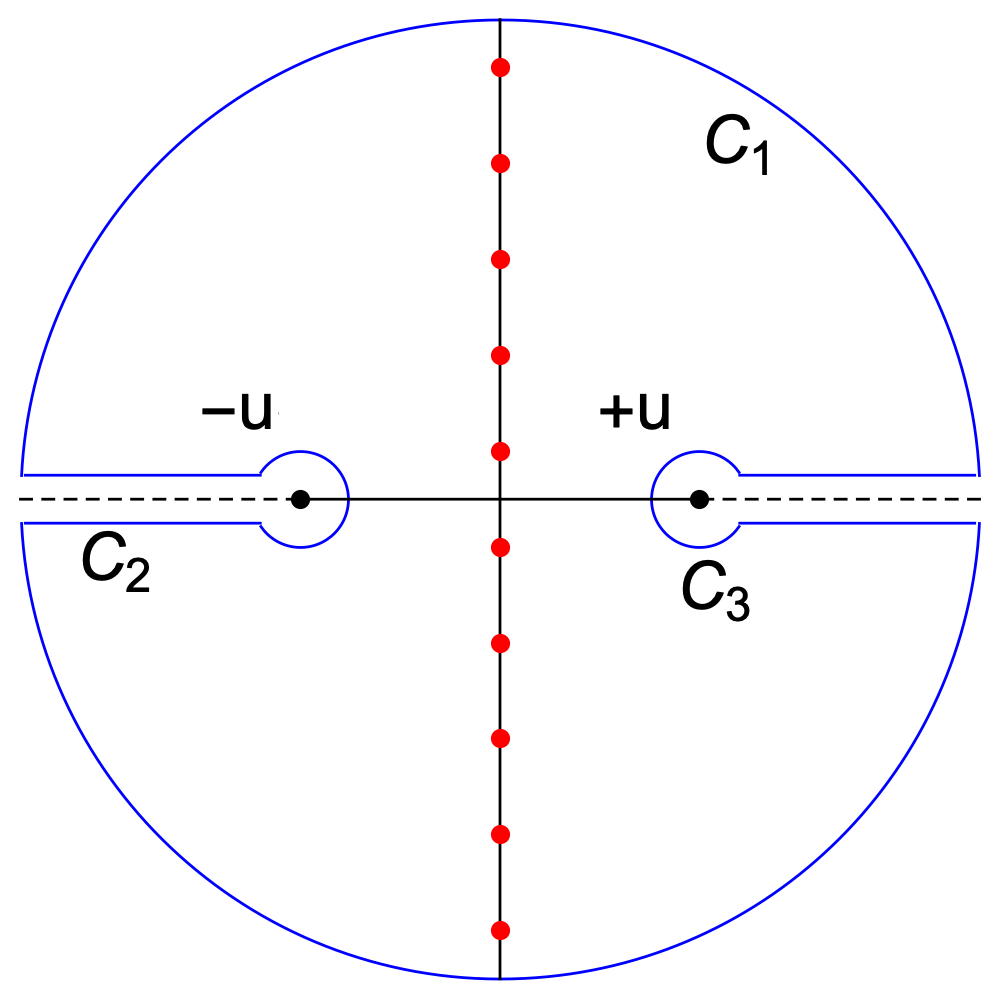}
\includegraphics[width=3in,height=3in]{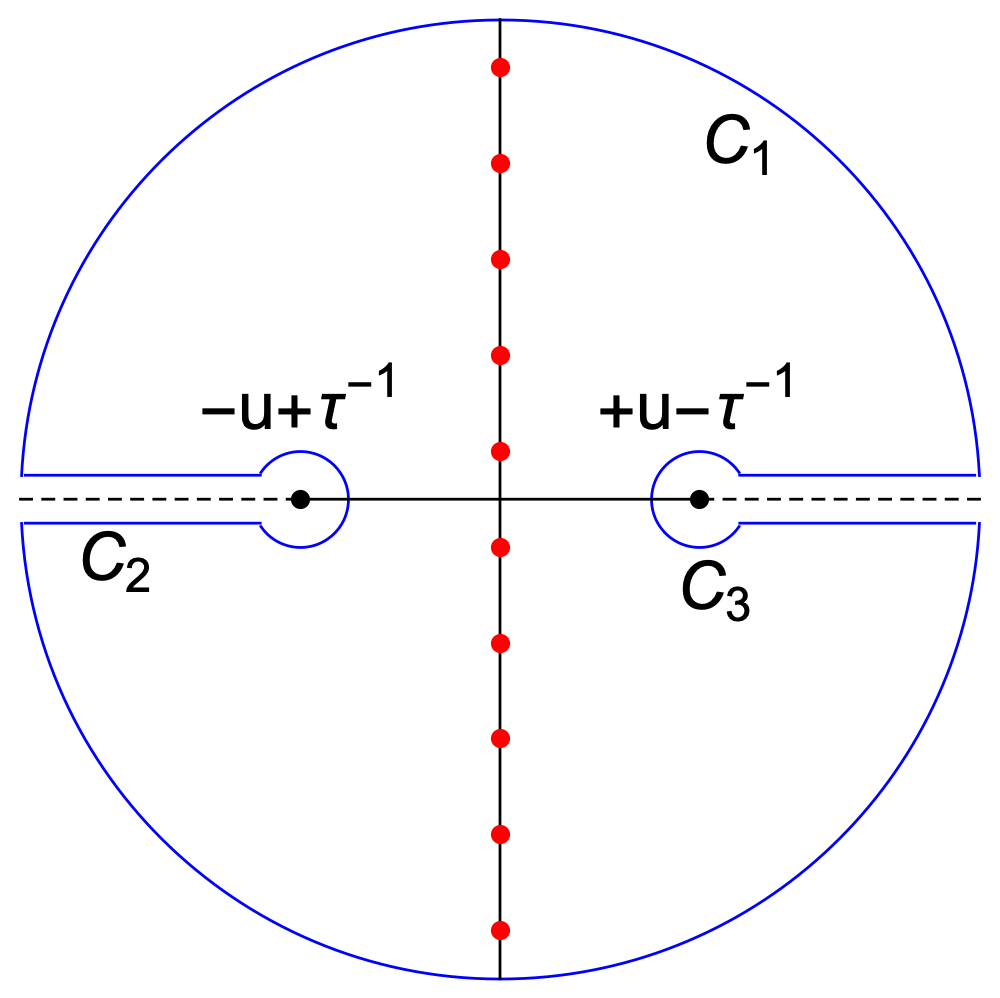}
\caption{Contour for evaluating Matsubara sums with branch points at $\pm u $ and $\pm (u - \tau^{-1})$ for the clean limit (left) and weak impurity scattering limit (right) respectively. The branch cuts are denoted by dashed lines and fermionic poles by red disks.} \label{Contour}
\end{figure}

In this Appendix, we evaluate the fractional Matsubara sums appearing in the main text. We recall that an expansion of the inverse fluctuation propagator in the parameter $r$ in the static limit gives
\beq 
L^{-1}(\bs q, \Omega\rightarrow 0) &\simeq& -g^{-1} + \Pi^{(0)}(0,0) + \Pi^{(2)}(\bs q,0), \\ \nonumber
\eeq
where $\Pi^{(0)}(0,0) = \frac{m}{4}\left(2 S_1 - u^2 S_3\right)$, $\Pi^{(2)}(\bs q,0) = -\frac{m r^2}{32}\left(2 S_3 - u^2 S_5\right)$ and  $S_{\nu} = \frac{1}{\beta} \sum_{\epsilon_n} (\epsilon_n^2 + u^2 )^{-\nu/2} $. We now wish to evaluate $S_{\nu}$ for odd $\nu = 1, 3, 5,..$. \\ \newline
\textit{Case 1}, $\nu=1$: We want to evaluate the divergent sum  $S_{1} = \frac{1}{\beta} \sum_{\epsilon_n} (\epsilon_n^2 + u^2 )^{-1/2} $. To this end, consider an integral over the contour $C$ in the complex plane (shown in Fig.~\ref{Contour} (left)) with branch points at $\pm u$ and a branch cut extending out to $\pm\infty$ from their respective branch points. Using Cauchy's theorem, we can relate this integral to the sum $S_1$ using the formula 
\beq
\oint_{C = C_1 + C_2 + C_3} \frac{g_F(z) dz}{(-z^2 + u^2)^{1/2}} = \frac{2\pi i}{\beta} \sum_{\epsilon_n} \frac{1}{(\epsilon_n^2 + u^2)^{1/2}},
\eeq
where $g_F(x) = \frac{1}{2} \tanh\left(\frac{\beta x}{2}\right)$, and the right hand side is simply a sum of residues of the poles at the fermionic Matsubara frequencies. 
 To determine this integral, we divide the total contour into three parts, $C_{1,2,3}$, and evaluate each individually. We begin with the circular contour $C_3$ with a radius ($\epsilon$) that has a zero limiting value. This is given as
 \beq
\frac{1}{2\pi i} \oint_{C_3} \frac{g_F(z) dz }{(-z^2 + u^2)^{1/2}}=- \frac{1}{2\pi} \oint_{C_3} \frac{g_F(z) dz}{(z+u)^{1/2}(z-u)^{1/2}}.
 \eeq
We can parameterize the variable near the $z=u$ branch point as $z = u + \epsilon e^{i\phi}$ where $0\leq \phi \leq 2\pi$ and $dz = i \epsilon e^{i\phi} d\phi$. With this substitution we obtain
 \beq
\frac{1}{2\pi i} \oint_{C_3} \frac{g_F(z) dz }{(-z^2 + u^2)^{1/2}}=- \frac{i \sqrt{\epsilon}}{2\pi} \int_0^{2\pi} \frac{g_F(u + \epsilon e^{i\phi}) e^{i\phi/2} d\phi}{(2 u + \epsilon e^{i\phi})^{1/2}},
 \eeq
leading to a vanishing contribution as $\sqrt{\epsilon}$ as $\epsilon \rightarrow 0$. We can similarly parameterize the variable near the $z=-u$ branch point as $z = -u + \epsilon e^{i\phi}$ where $-\pi \leq \phi \leq \pi$ and $dz = i \epsilon e^{i\phi} d\phi$. This contribution to the total integral also vanishes as $\sqrt{\epsilon}$ as $\epsilon \rightarrow 0$. We now consider the contour integral over the large circle $C_1$ with radius $R$. As the circle is centered around $z=0$, we can use the parameterization $z = R e^{i\phi}$ where $0\leq \phi \leq 2\pi$ and $dz = i R e^{i\phi} d\phi$. With this substitution, the $C_1$ contribution is 
 \beq
\frac{1}{2\pi i} \oint_{C_1} \frac{g_F(z) dz }{(-z^2 + u^2)^{1/2}}= \frac{-1}{2\pi} \int_0^{2\pi}  \frac{d\phi (i R e^{i\phi}) g_F(R e^{i\phi})}{(R^2 e^{2i\phi} - u^2)^{1/2}}.
 \eeq
 Taking the limit $R\rightarrow \infty$, we have 
  \beq
\frac{1}{2\pi i} \oint_{C_1} \frac{g_F(z) dz }{(-z^2 + u^2)^{1/2}} \rightarrow - \frac{i}{2\pi} \int_0^{2\pi} d\phi~g_F(R e^{i\phi}) = \#,
 \eeq
 where $\#$ is a constant independent of the physical parameters $u$ and $T$ as $R\rightarrow \infty$. Finally, we consider the contribution from the contour $C_2$ (which we denote as $I_{C_2}$) formed by the straight lines originating from the branch points $\pm u$ which is given by
  \beq
I_{C_2} = \frac{1}{2\pi i} \oint_{C_2} \frac{g_F(z) dz }{(-z^2 + u^2)^{1/2}}=- \frac{1}{2\pi} \oint_{C_2} \frac{g_F(z) dz}{(z+u)^{1/2}(z-u)^{1/2}}.
 \eeq
$I_{C_2}$ can be split into four individual contributions depending on whether the contour is in the upper/lower complex plane or positive/negative real axis. Denoting $z_{\pm}$ as the variable in the upper/lower complex plane we can write 
 \beq \nonumber
 -I_{C_2} &=& \frac{1}{2\pi} \int_{u+\epsilon}^{\infty} \frac{g_F(z_-) dz_-}{(z_- + u)^{1/2}(z_- - u)^{1/2}} + \frac{1}{2\pi} \int_{\infty}^{u+\epsilon} \frac{g_F(z_+) dz_+}{(z_+ + u)^{1/2}(z_+ - u)^{1/2}} \\
 && +  \frac{1}{2\pi} \int_{-u-\epsilon}^{-\infty} \frac{g_F(z_+) dz_+}{(z_+ + u)^{1/2}(z_+ - u)^{1/2}}  + \frac{1}{2\pi} \int_{-\infty}^{-u-\epsilon} \frac{g_F(z_-) dz_-}{(z_- + u)^{1/2}(z_- - u)^{1/2}}.
 \eeq
 Since $(z\pm u)^{1/2}$ and $g_F(z)$ are analytic across the branch points $z = \pm u$ respectively, we can rewrite $I_{C_2}$ as 
 \beq \nonumber
 -I_{C_2}  &=& \frac{1}{2\pi}  \int_{u+\epsilon}^{\infty} \frac{g_F(z) dz}{(z+ u)^{1/2}} \left[ \frac{1}{(z_- - u)^{1/2}} - \frac{1}{(z_+ - u)^{1/2}} \right] \\
 && +  \frac{1}{2\pi}  \int^{-u-\epsilon}_{-\infty} \frac{g_F(z) dz}{(z - u)^{1/2}} \left[ \frac{1}{(z_- + u)^{1/2}} - \frac{1}{(z_+ + u)^{1/2}} \right]. 
 \eeq
 The quantities in the brackets above can be evaluated using the relations
  \beq \label{DiscontinuityRelation1}
 \frac{1}{(z_+ - u)^{1-\alpha}} - \frac{1}{(z_- - u)^{1-\alpha}} &=& \frac{2 i \sin \pi \alpha~e^{- i \pi(1-\alpha)}}{|z-u|^{1-\alpha}}\\  \label{DiscontinuityRelation2}
   \frac{1}{(z_+ + u)^{1-\alpha}} - \frac{1}{(z_- + u)^{1-\alpha}} &=&  \frac{-2 i \sin \pi \alpha}{|z+u|^{1-\alpha}}.
 \eeq
Using these relations by setting $\alpha \rightarrow 1/2$ we can simplify $I_{C_2}$ to write
\beq
I_{C_2} = \lim_{\epsilon \rightarrow 0} \frac{-1}{\pi} \int_{u+\epsilon}^{\infty} \frac{\left(g_F(-z) - g_F(z)\right) dz}{\sqrt{z^2 - u^2}} = \text{P.V} \left [\frac{1}{\pi} \int_{u}^{\infty}\frac{\tanh\left(\frac{\beta z}{2}\right) dz}{\sqrt{z^2 - u^2}} \right], 
\eeq
 where $\text{P.V}$ denotes principal value. As is evident from the form above, $I_{C_2}$ (and consequenty $S_1$) is UV divergent; hence, we set a cut-off energy parameter $\Lambda$ to isolate the divergence. Changing variables $z = z' u$, we can evaluate the integral in the strong coupling limit $\beta u \gg 1$ where we can approximate $\tanh x \simeq 1 - 2 e^{-2 x}$. Taking the limit $\Lambda / u \gg 1$ and substituting $I_{C_2}$ back into $S_1$ we have
\beq
S_1 \simeq \frac{1}{\pi} \ln\left(\frac{\Lambda}{u}\right) - \frac{2}{\pi} K(0, \beta u) + \#,
\eeq
where $K(x,y)$ is the modified Bessel function of the second kind. \newline \\
\textit{Case 2}, $\nu=3, 5, ..$~: We will now evaluate the convergent sums $S_{\nu} = \frac{1}{\beta} \sum_{\epsilon_n} (\epsilon_n^2 + u^2)^{-\nu/2}$ where $\nu = 3, 5, ..$. Similar to the case of $\nu =1$, we can break up the sums into three individual pieces of integration around $C_{1,2,3}$ shown in Fig.~\ref{Contour} (left). Hence we write 
\beq
\left[\oint_{C_1} +\oint_{C_2} + \oint_{C_3} \right]\frac{g_F(z) dz}{(-z^2 + u^2)^{\nu/2}} = \frac{2\pi i}{\beta} \sum_{\epsilon_n} \frac{1}{(\epsilon_n^2 + u^2)^{\nu/2}}.
\eeq
We begin evaluating the large contour $C_1$ by replacing $z= R e^{i\phi}$ and $dz = i R e^{i\phi} d\phi$. With this substitution we have
\beq
\frac{1}{2 \pi i} \oint_{C_1} \frac{g_F(z) dz}{(-z^2 + u^2)^{\nu/2}} = \frac{e^{-i \pi \nu/2}}{2\pi i} \int_0^{2\pi} \frac{g_F(R e^{i\phi}) (i R e^{i\phi})d\phi}{(R^2 e^{2i\phi} - u^2)^{\nu/2}}.
\eeq
Taking the limit $R\rightarrow \infty$ the integral becomes
\beq
\frac{1}{2 \pi i} \oint_{C_1} \frac{g_F(z) dz}{(-z^2 + u^2)^{\nu/2}} = \frac{e^{-i \pi \nu/2}}{2\pi i} \int_0^{2\pi} \frac{g_F(R e^{i\phi}) (i R e^{i\phi})d\phi}{R^{\nu} e^{\nu i \phi}} \sim \frac{1}{R^{\nu -1}} \rightarrow 0 ~~~~~\text{for} ~~~~\nu>1.
\eeq
Hence the contour $C_1$ does not contribute to $S_{\nu}$. \newline \\ We will now show that the IR divergent contribution from contour $C_3$ is cancelled with that of $C_2$ yielding an $S_\nu$ that is finite as must be anticipated for $\nu=3, 5, ..$. We begin with the $C_3$ contribution from the $z= u$ branch point. Like before, we make the substitution $ z = u + \epsilon e^{i\phi}$ where $0\leq \phi \leq 2\pi$ and we obtain
\beq
\frac{1}{2 \pi i} \oint_{C_3, z=u} \frac{g_F(z) dz}{(-z^2 + u^2)^{\nu/2}} = \frac{e^{-i \pi \nu/2}}{2\pi i} \int_0^{2\pi} \frac{g_F(u+ \epsilon e^{i\phi}) (i\epsilon e^{i\phi}) d\phi}{(2 u + \epsilon e^{i\phi})^{\nu/2}(\epsilon e^{i\phi})^{\nu/2}},~~~~~~~~~\text{for}~~~~~~~~~z=u.
\eeq
Taking the limit of $\epsilon \rightarrow 0$ and solving the $\phi$ integral we have the IR divergent term from $z=u$
\beq
\frac{1}{2 \pi i} \oint_{C_3, z=u} \frac{g_F(z) dz}{(-z^2 + u^2)^{\nu/2}} = -\frac{e^{-i \pi \nu/2}}{(2 u)^{\nu/2}\pi i}\frac{g_F(u)}{\epsilon^{\frac{\nu}{2} - 1}}\left[\frac{2}{2-\nu}\right] ~~~~~~~~ \text{for}~~~~~~~z=u,~~~~~ \nu=3,5,..
\eeq
Similarly the contribution from the $z=-u$ branch point can be obtained by the substitution $z = -u + \epsilon e^{i\phi}$ where $-\pi\leq\phi\leq\pi$. The result is equal to that obtained for the $z=u$ case discussed above and thus gives a total contribution from the $C_3$ contour
\beq \label{IRDivergence}
\frac{1}{2 \pi i} \oint_{C_3} \frac{g_F(z) dz}{(-z^2 + u^2)^{\nu/2}} = -\frac{2e^{-i \pi \nu/2}}{(2 u)^{\nu/2}\pi i}\frac{g_F(u)}{\epsilon^{\frac{\nu}{2} - 1}}\left[\frac{2}{2-\nu}\right] ~~~~~~~~~ \nu=3,5,..
\eeq
This term is IR divergent as $\sim \frac{1}{\epsilon^{\frac{\nu}{2} -1}}$.  We now evaluate the contribution from the $C_2$ contour by following a similar procedure as the $\nu =1 $ case. We have
\beq \nonumber
\frac{1}{2\pi i}\oint_{C_2} \frac{e^{-i\pi\nu/2} g_F(z) dz}{(z^2 - u^2)^{\nu/2}} &=& \frac{e^{-i\pi \nu/2}}{2\pi i}\Bigg\{ \int_{u+\epsilon}^{\Lambda} \frac{g_F(z) dz}{(z+ u)^{\nu/2}} \Big[ \frac{1}{ (z_- - u)^{\nu/2}}  -   \frac{1}{ (z_+ - u)^{\nu/2}}   \Big] \\
&&+ \int^{-u-\epsilon}_{-\Lambda} \frac{g_F(z) dz}{(z- u)^{\nu/2}} \Big[ \frac{1}{ (z_- + u)^{\nu/2}}  -   \frac{1}{ (z_+ + u)^{\nu/2}}   \Big] \Bigg\}.
\eeq
We can now utilize Eqs.~\ref{DiscontinuityRelation1} and~\ref{DiscontinuityRelation2} to substitute for quantities appearing in the square brackets above. We make the replacement $\alpha \rightarrow 1 - \frac{\nu}{2}$ and after simplifications we are left with
\beq \nonumber
\frac{1}{2\pi i}\oint_{C_2} \frac{e^{-i\pi\nu/2} g_F(z) dz}{(z^2 - u^2)^{\nu/2}} &=&\frac{e^{-i\pi \nu/2}}{2\pi i}\Bigg\{ \int_{u+\epsilon}^{\Lambda} \frac{g_F(z) dz}{(z+ u)^{\nu/2}} \Big[ \frac{-2 i \sin \pi(1-\frac{\nu}{2})e^{-i\pi\nu/2}}{|z-u|^{\nu/2}}\Big] \\
&&+ \int^{-u-\epsilon}_{-\Lambda} \frac{g_F(z) dz}{(z- u)^{\nu/2}} \Big[ \frac{2 i \sin \pi(1-\frac{\nu}{2})}{|z+u|^{\nu/2}}\Big] \Bigg\} \\
&=& -\frac{e^{-i\pi \nu}}{2\pi i} \frac{2 i \sin \pi (1- \frac{\nu}{2})}{u^{\nu-1}} \int_{1 + \frac{\epsilon}{u}}^{\Lambda/u} \frac{dz' \tanh\left(\frac{\beta u z'}{2}\right)}{|z'^2 - 1|^{\nu/2}}~~~~~~~~~~~~\nu = 3, 5,... 
\eeq
where in the last step we changed variables $z=  u z'$. To be able to solve the integrals above and extract the IR divergence, we perform the strong coupling expansion $\tanh x \simeq 1 - 2 e^{-2x}$. The integral of the first term in the expansion gives for $\Lambda/u \rightarrow \infty$
\beq
\int_{1+ \frac{\epsilon}{u}}^{\infty} \frac{dz'}{|z'^2 -1|^{\nu/2}} = \left[ \frac{\sqrt{\pi}\Gamma\left(\frac{\nu-1}{2}\right)\sin\left(\frac{\pi \nu}{2}\right)}{2~\Gamma(\nu/2)} + \frac{(2~\epsilon)^{1-\nu/2}}{(\nu-2)u^{1-\nu/2}} \right]~~~~~~~~~~~~\nu = 3, 5,... 
\eeq
where the second term diverges as $\sim \frac{1}{\epsilon^{\frac{\nu}{2} -1}}$ and cancels the IR divergence in Eq.~\ref{IRDivergence} for $\beta u \rightarrow \infty$. Therefore, we only need to keep the principal value of the integral over the contour $C_2$, i.e., 
\beq 
\frac{1}{2\pi i}\oint_{C_2} \frac{e^{-i\pi\nu/2} g_F(z) dz}{(z^2 - u^2)^{\nu/2}} &\simeq& \frac{e^{-i\pi \nu}}{2\pi i} \frac{2i \sin \pi\left(1 - \nu/2\right)}{u^{\nu-1}} \text{P.V} \left\{\int_1^{\infty} dz' \frac{\left(1- 2 e^{-\beta u z'}\right)}{|z'^2 -1|^{\nu/2}}\right\} ~~~~~~\nu = 3, 5, ...
\eeq
The principal value integral can be solved exactly and can be combined with the $\nu=1$ case to give the sum $S_{\nu}$ as
\begin{align}
S_{\nu} =\begin{cases}
    \text{$\frac{1}{\pi} \ln\frac{\Lambda}{u} - \frac{2}{\pi} K(0,\kappa)$}~~~~~~~~~~~~~~~~~~~~~~~~~~~~~~~~~~~~~~~~~~~~~~~~~~~~~~~\text{for $\nu=1$ }\\
       \text{$\frac{e^{- i \pi \nu} \Gamma\left(1- \frac{\nu}{2}\right) \sin \pi \left( 1- \frac{\nu}{2} \right)}{u^{\nu-1}2 \pi^{3/2}}\left[ 2^{\frac{5-\nu}{2}} \kappa^{\frac{\nu -1}{2}} K\left( \frac{\nu-1}{2}, \kappa\right) - \Gamma\left(\frac{\nu -1}{2}\right) \right]$}~~~~~~\text{for $\nu = 3, 5, ..$.} 
     \end{cases} 
\end{align}
where $\kappa \equiv \beta u$ and $K(x,y)$ is the modified Bessel function of the second kind. This is Eq.~\ref{Minimum} in the main text. 
\section{Appendix D}
In this section, we will evaluate relevant Matsubara sums to arrive at the expression for the fluctuation propagator in the weak coupling ($\kappa\ll1$) clean limit ($\tau \rightarrow \infty$). We begin with the small $u$ expansions of the pair susceptibilities appearing in the main text (the powers $(0)$ and $(2)$ on top of the pair susceptibility components denote powers of the small $r\equiv \frac{p_f q}{m}$  expansion)
\beq 
\Pi^{(0)}(0, 0)&\simeq& \frac{m}{4\beta} \sum_{\epsilon_n} \left[ \frac{2}{|\epsilon_n|} - \frac{2 u^2}{|\epsilon_n|^3} \right] \\ 
 \Pi^{(2)}(\bs q,0) &\simeq& \frac{-m r^2}{32 \beta}\sum_{\epsilon_n} \left[ \frac{2}{|\epsilon_n|^3} - \frac{4 u^2}{|\epsilon_n|^5}\right].
\eeq 
Consider the sum
\beq
\sum_{n=-\infty}^{n=\infty} \frac{1}{|n + \frac{1}{2} + x|^p} = \left( \sum_{n=0}^{\infty}  + \sum_{n=-\infty}^{-1}\right)\frac{1}{|n + \frac{1}{2} + x|^p}~~~~~~p=1, 2, 3,..
\eeq
Inverting signs of the summation variable in the second term, then making the variable shift $n'=n-1$ and combining terms we get
\beq
\sum_{n=-\infty}^{n=\infty} \frac{1}{|n + \frac{1}{2} + x|^p} = \sum_{n=0}^{\infty}\left(\frac{1}{|n + \frac{1}{2} + x|^p} + \frac{1}{|n + \frac{1}{2} - x|^p}\right).
\eeq
Using this relation we can write $\Pi^{(0)}(0, 0)$ as (for $x=0$)
\beq
\Pi^{(0)}(0, 0)&\simeq&\frac{m}{4\beta} \sum_{n=0}^{\infty}\left[ \frac{4}{2\pi T\left( n + \frac{1}{2}\right)} - \frac{4 u^2}{ (2\pi T)^3 \left( n + \frac{1}{2}\right)^3} \right].
\eeq
Noting that $\sum_{n=0}^{\Lambda/2\pi T} (n+ \frac{1}{2})^{-1} \simeq \ln\left(\frac{\Lambda}{2\pi T}\right) - \psi(1/2)$ and $\sum_{n=0}^{\infty} (n+ \frac{1}{2})^{-3} = -\frac{1}{2} \psi''(1/2)$, we arrive at
\beq
\Pi^{(0)}(0, 0)&\simeq& N_0\left[  \ln\left(\frac{\Lambda}{2\pi T}\right) - \psi(1/2) + \frac{u^2}{2(2\pi T)^2} \psi''(1/2)  \right].
\eeq
Similarly we can write
\beq
 \Pi^{(2)}(\bs q,0) &\simeq& \frac{-mr^2}{32\beta} \sum_{n=0}^{\infty} \left[ \frac{4}{(2\pi T)^3 \left( n + \frac{1}{2}\right)^3}  - \frac{8 u^2}{(2\pi T)^5 \left( n + \frac{1}{2}\right)^5}\right].
\eeq
Noting again that $\sum_{n=0}^{\infty} (n+ \frac{1}{2})^{-5} = -\frac{1}{24} \psi^{(4)}(1/2)$, we arrive at
\beq
 \Pi^{(2)}(\bs q,0) &\simeq& \frac{-N_0 r^2}{32} \left[ \frac{2 |\psi''(1/2)|}{(2\pi T)^2} -  \frac{u^2 |\psi^{(4)}(1/2)|}{3(2\pi T)^4}\right].
\eeq
Combining $\Pi^{(0)}(0, 0)$ and $ \Pi^{(2)}(\bs q,0)$ we obtain the fluctuation propagator in the weak coupling, clean limit as
\beq
L^{-1}(\bs q, \Omega\rightarrow0) &=& -\frac{1}{g} + N_0 \left[  \ln~\frac{\Lambda}{2\pi T} - \psi\left(\frac{1}{2}\right) - \frac{u^2 C_2}{8\pi^2T^2}\right] -\frac{N_0 r^2}{128 \pi^2 T^2} \left[ 2 C_2- \frac{u^2 C_4}{12 \pi^2 T^2} \right].
\eeq

\section{Appendix E}
In this Appendix we derive the fluctuation propagator in the weak coupling limit ($\kappa \ll 1$) with dilute impurities ($\theta\equiv T\tau\gg1$). We recall the static long-wavelength pair susceptibility for weak impurity scattering from the main text 
\beq
\Pi(\bs q \rightarrow 0, \Omega =0) &=& \frac{1}{\beta}\sum_{n} \frac{2\pi N_0 \tau \tilde{A}(\tilde{\epsilon}_n, u)}{2\pi N_0 \tau - \tilde{A}(\tilde{\epsilon}_n, u)}\\
\tilde{A}(\tilde{\epsilon}_n, u) &=& \frac{2 \pi N_0}{4}\left[ \frac{2}{\tilde{\epsilon}_n'} - \frac{u^2}{\tilde{\epsilon}_n'^3}\right].
\eeq  
Keeping only terms quadratic in $u$, we obtain
\beq
\Pi(\bs q \rightarrow 0, \Omega =0) &\simeq& \frac{2\pi N_0}{\beta} \sum_{n=-\infty}^{\infty} \left[ \frac{\tau}{2|\epsilon_n + \frac{\text{sgn}(\epsilon_n)}{2\tau}|\tau -1} - \frac{2 u^2 \tau^2}{\left(2|\epsilon_n + \frac{\text{sgn}(\epsilon_n)}{2\tau}|\tau -1\right)^2 |\epsilon_n + \frac{\text{sgn}(\epsilon_n)}{2\tau}|}\right] \\
&=&\frac{4\pi N_0}{\beta} \sum_{n=0}^{\infty}\left[ \frac{1}{2\epsilon_n} - \frac{u^2}{2\left(\epsilon_n + \frac{1}{2\tau} \right)\epsilon_n^2}\right] \\
&=&N_0  \sum_{n=0}^{\infty} \left[ \frac{1}{\left(n + \frac{1}{2}\right)} - 4 u^2 \tau^2 \left\{ \frac{1}{\left(n + \frac{1}{2} + \frac{1}{4\pi T \tau}\right)} + \frac{1}{4\pi T \tau\left( n + \frac{1}{2} \right)^2}  - \frac{1}{\left(n + \frac{1}{2} \right)}\right\} \right].
\eeq
In the second line we changed the summation to positive integers by inverting sign of the summation variable. Using definitions and properties of Gamma functions we can write the final expression for the pair susceptibility ($\theta \equiv T\tau$)
\beq
\Pi(\bs q \rightarrow 0, \Omega =0) = N_0\left[ \ln~\left(\frac{\Lambda}{4\pi T}\right) - \psi\left(\frac{1}{2}\right) + 4 u^2 \tau^2 \left[  \psi\left(\frac{1}{2} + \frac{1}{4\pi \theta}\right)  -\frac{1}{4\pi\theta} \psi'(\frac{1}{2}) - \psi\left(\frac{1}{2}\right)\right]\right],
 \eeq
which appears in the final expression of the fluctuation propagator in the main text.

\section{Appendix F}
In this Appendix, we evaluate Matsubara sums appearing in the strong coupling ($\kappa \gg 1$), dilute impurity ($\theta \gg 1$) limit. The derivation follows along similar lines as the case of the clean limit but with branch points shifted by $\tau^{-1}$. See Fig.~\ref{Contour} (right) for a sketch of the integration contour chosen. We begin by recalling the pair susceptibility in the limit $\kappa \gg 1$, $\theta \gg 1$ where vertex corrections can be ignored,
\beq
\Pi(\bs q, \Omega_k) \simeq \frac{1}{\beta} \sum_n P(\bs q, \tilde{\epsilon}_{n+k}, -\tilde{\epsilon}_n).
\eeq
For $\Omega_k =0$, one can perform an expansion in $r$ similar to the clean case and write 
\beq \nonumber
P(\bs q, \tilde{\epsilon}_n, -\tilde{\epsilon}_n) &\simeq& P^{(0)}(\bs q =0, \tilde{\epsilon}_n, -\tilde{\epsilon}_n) + P^{(2)}(\bs q, \tilde{\epsilon}_n, -\tilde{\epsilon}_n), \\
 P^{(0)}(0,0) &=& \frac{\beta m}{4}\left(2 \tilde{S}_1 - u^2 \tilde{S}_3\right), \\
  P^{(2)}(\bs q,0) &=& -\frac{\beta m r^2}{32}\left(2 \tilde{S}_3 - u^2 \tilde{S}_5\right)
\eeq
 and  $\tilde{S}_{\nu} = \frac{1}{\beta} \sum_{\epsilon_n} (\tilde{\epsilon}_n^2 + u^2 )^{-\nu/2} $. To evaluate $\tilde{S}_{\nu} $, consider the integral over the contour $C$ in Fig.~\ref{Contour} (right)
 \beq
 \tilde{I}_{\nu} = \oint_C \frac{g_F(z) dz}{ \left[u^2 - \left(z + i \frac{\text{sgn}(z/i)}{2 \tau}\right)^2\right]^{\nu/2}} = \oint_C \frac{e^{- i \pi\nu/2}g_F(z) dz}{ \left[-u^2 + z^2\left(1 +  \frac{1}{2 \tau |z|}\right)^2\right]^{\nu/2}},
 \eeq
 where we used the definition of the complex signum function $\text{sgn}(z) = z/|z|$ to obtain the right hand side. For $u> \frac{1}{2 \tau}$, the branch points can be solved as $z = \pm\left( u - \frac{1}{2\tau}\right)$ with the branch cuts originating from these points to $\pm \infty$ (see Fig.~\ref{Contour} (right)). Using Cauchy's theorem, we can easily see that the sum $\tilde{S}_{\nu} = \frac{1}{2\pi i} \tilde{I}_{\nu}$. Like the clean case, in the limit $\kappa\gg 1$, $\theta \gg1$ and $u> \frac{1}{2\tau}$, the non-trivial contribution to the summation comes from the $C_2$ part of the contour. This is true for both the $\nu =1$ and $\nu = 3, 5,...$ cases. Extending the results for the clean case, we obtain for $\tau< \infty$ and $\theta \gg 1$
 \beq
 \tilde{S}_{\nu} = \frac{e^{-i\pi \nu} \sin \pi\left(1 - \frac{\nu}{2}\right)}{\pi} \int_{u - \frac{1}{2\tau} + \epsilon}^{\Lambda} dz\frac{\left(g_F(z) - g_F(-z)\right)}{ \left[-u^2 + z^2\left(1 +  \frac{1}{2 \tau |z|}\right)^2\right]^{\nu/2}},
 \eeq
 where $\Lambda$ can be extended to infinity for the cases $\nu  = 3, 5, ..$. As the integration is now over real variables, we can make the substitution $z = xu$ to yield
 \beq
  \tilde{S}_{\nu} = \frac{e^{-i\pi \nu} \sin \pi\left(1 - \frac{\nu}{2}\right)}{\pi u^{\nu -1}} \int_{1- \frac{1}{2 u \tau} + \frac{\epsilon}{u}}^{\Lambda/u} dx\frac{\tanh\left(\frac{\beta u x}{2}\right)}{\left[ \left( x + \frac{1}{2 u \tau}\right)^2 - 1\right]^{\frac{\nu}{2}}}.
 \eeq
 Setting $x+ \frac{1}{2 u \tau} = t$ and expanding the numerator hyperbolic function
 \beq
   \tilde{S}_{\nu} = \frac{e^{-i\pi \nu} \sin \pi\left(1 - \frac{\nu}{2}\right)}{\pi u^{\nu -1}} \int_{1 + \frac{\epsilon}{u}}^{\frac{\Lambda}{u} + \frac{1}{2 u \tau}} dt \frac{\left[\tanh\left(\frac{\beta u t}{2}\right) - \tanh\left(\frac{1}{4\pi \tau}\right) \right]}{\left(t^2 -1\right)^{\nu/2} \left[1- \tanh\left(\frac{\beta u t}{2}\right) \tanh\left(\frac{1}{4\pi \tau}\right) \right]}.
 \eeq
 Since $\beta u \gg 1$, $T\tau \gg 1$ we can rewrite the sum as
 \beq
    \tilde{S}_{\nu} \simeq \frac{e^{-i\pi \nu} \sin \pi\left(1 - \frac{\nu}{2}\right)}{\pi u^{\nu -1}}\text{P.V}\left[ \int_1^{\Lambda}dt \frac{1 - 2 e^{-\beta u t} e^{\frac{1}{2T\tau}}}{(t^2 - 1)^{\frac{\nu}{2}}} \right].
 \eeq
 The above integral looks similar to the one that is obtained in the clean limit except for the additional factor $e^{\frac{1}{2T\tau}}$. Hence, the expression for the fluctuation propagator in the strong coupling limit with dilute impurities can be readily generalized as
 \beq \nonumber
L^{-1}(\bs q, \Omega\rightarrow 0) &\simeq& - g^{-1} + N_0 \left[\ln\frac{\Lambda}{u} + \sqrt{\frac{\pi \kappa}{2}} e^{-\kappa}\right] \\
&& - \frac{N_0 r^2}{12 u^2} \left[ 1 + \sqrt{\frac{\pi \kappa^3}{8}} e^{-\kappa} \right].
\eeq
 This is the final expression that appears in the main text.

\section{Appendix G}
\begin{figure}[h!]
\includegraphics[width=2.7in,height=1.8in]{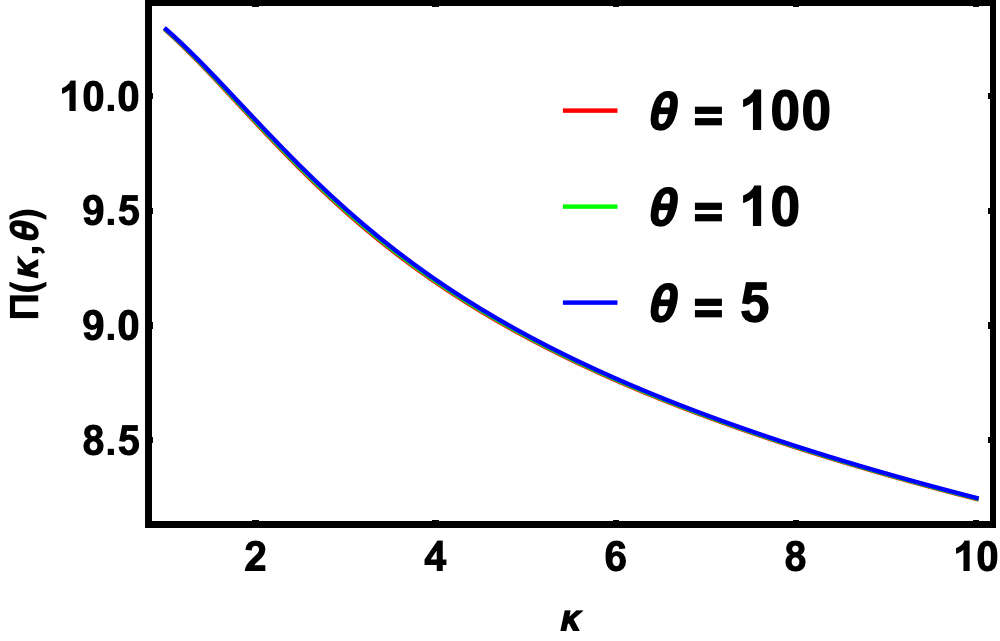}
\includegraphics[width=2.7in,height=1.8in]{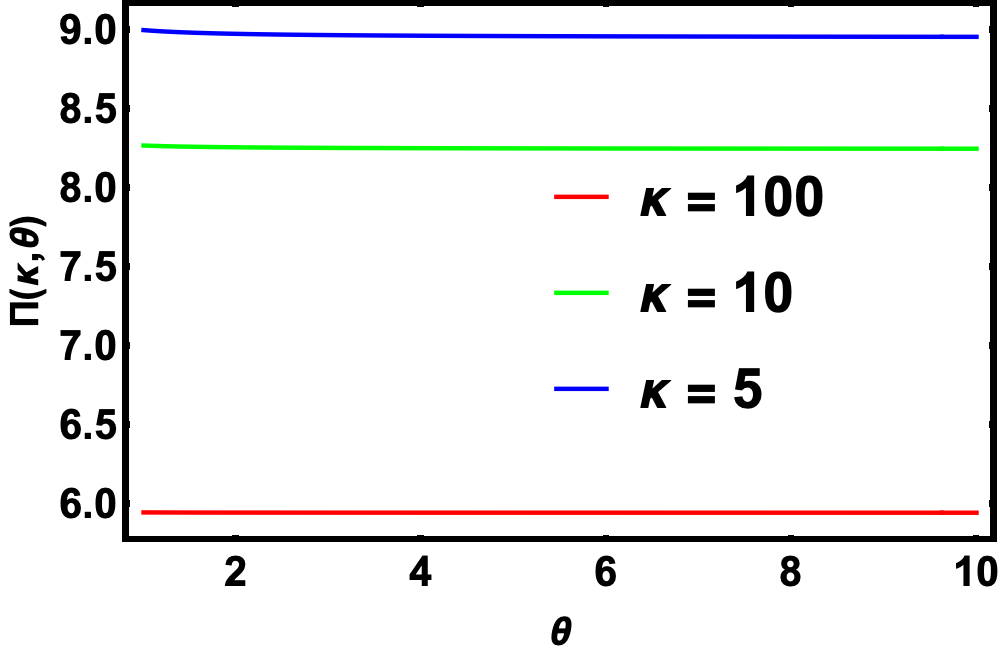}
\includegraphics[width=2.7in,height=1.8in]{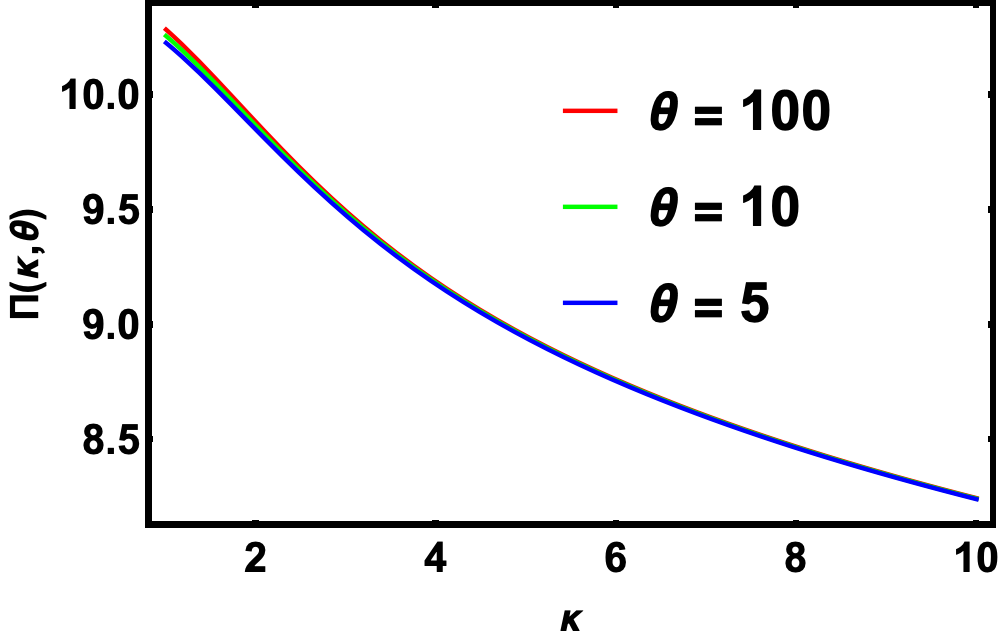}
\includegraphics[width=2.7in,height=1.8in]{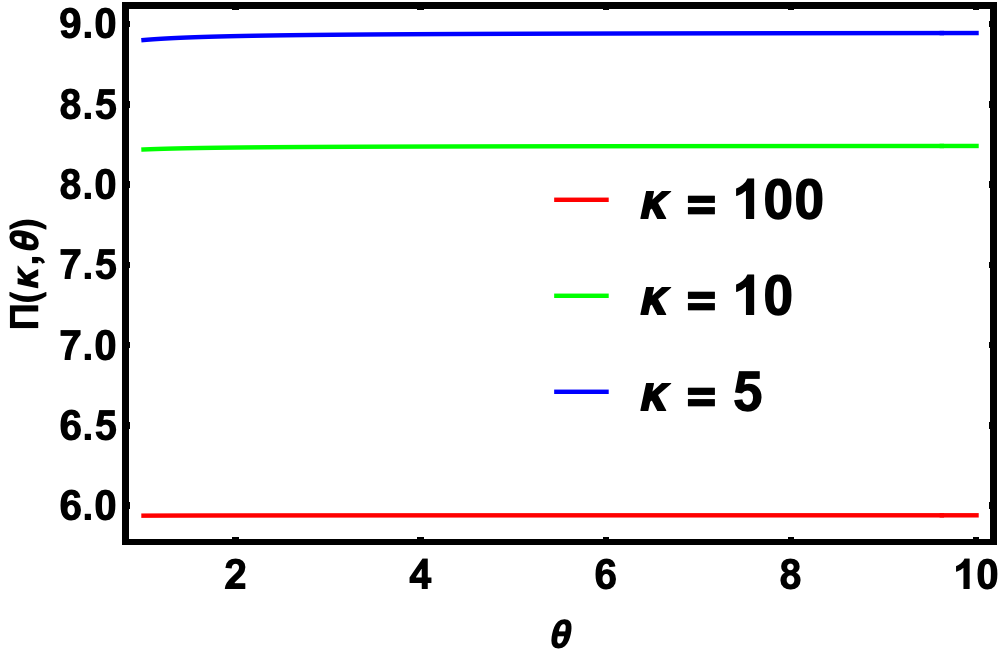}
\caption{Numerical plots of the pair susceptibility as a function of $\kappa = \beta u$ (left column) and $\theta = T\tau$ (right column) for $\kappa \gg 1$ and $\theta \gg 1$. The top (bottom) row are calculations with (without) vertex corrections due to impurities. The relatively flat behavior of the pair susceptibility as a function $\theta$ for $\kappa\gg1$ is due to the weak exponential dependence $\sim e^{\frac{1}{2\theta}}$. These results demonstrate that in the limit $\kappa\gg1$ and $\theta \gg1$, impurity vertex corrections have a negligible effect on the pair susceptibility.   } \label{LargeKLargeT}
\end{figure}
\begin{figure}[h!]
\includegraphics[width=3in,height=2in]{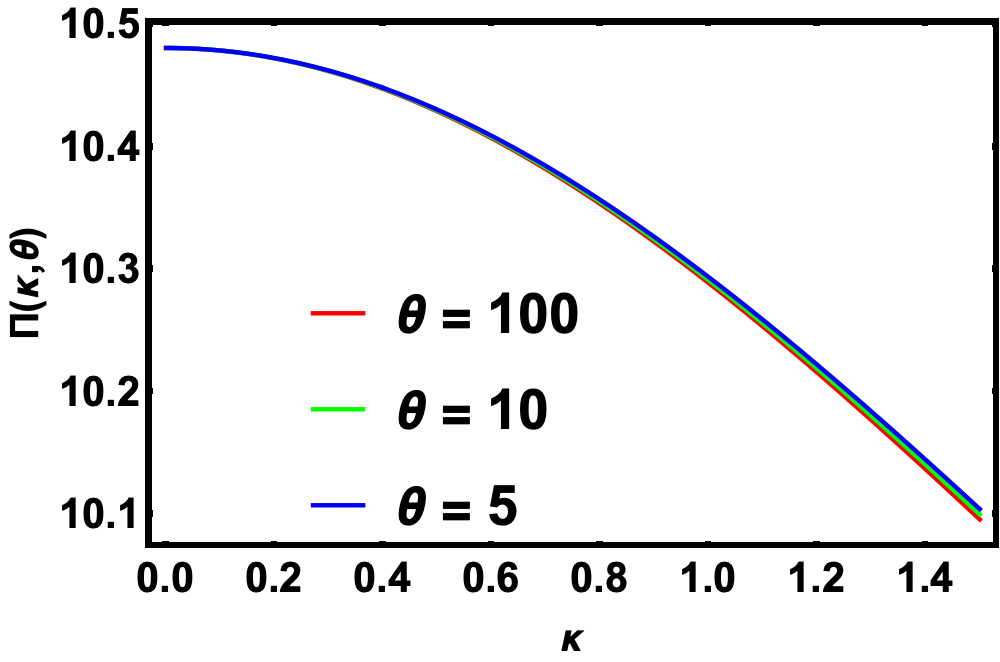}
\includegraphics[width=3in,height=2in]{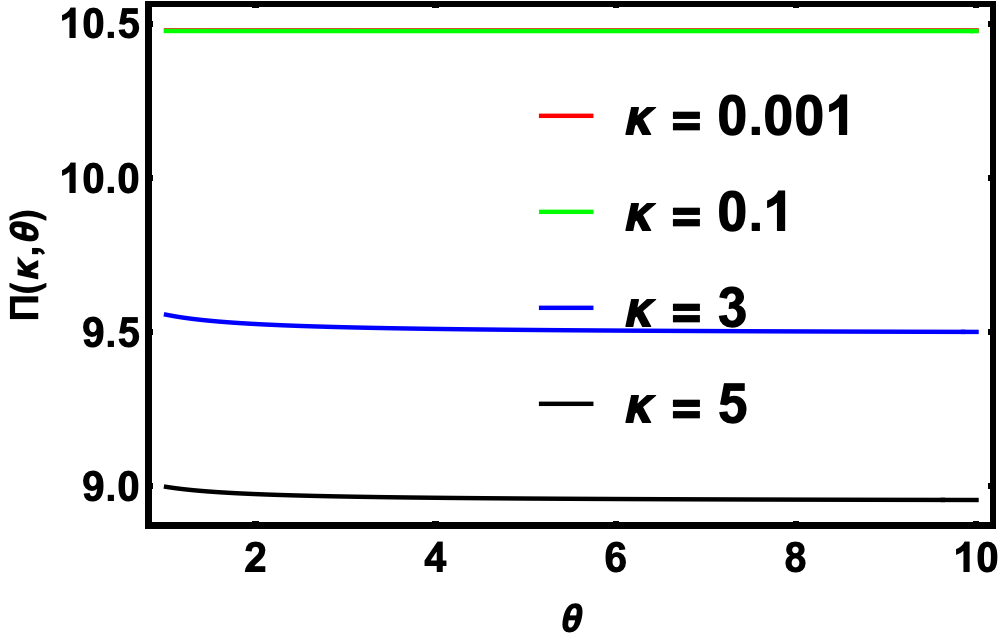}
\includegraphics[width=3in,height=2in]{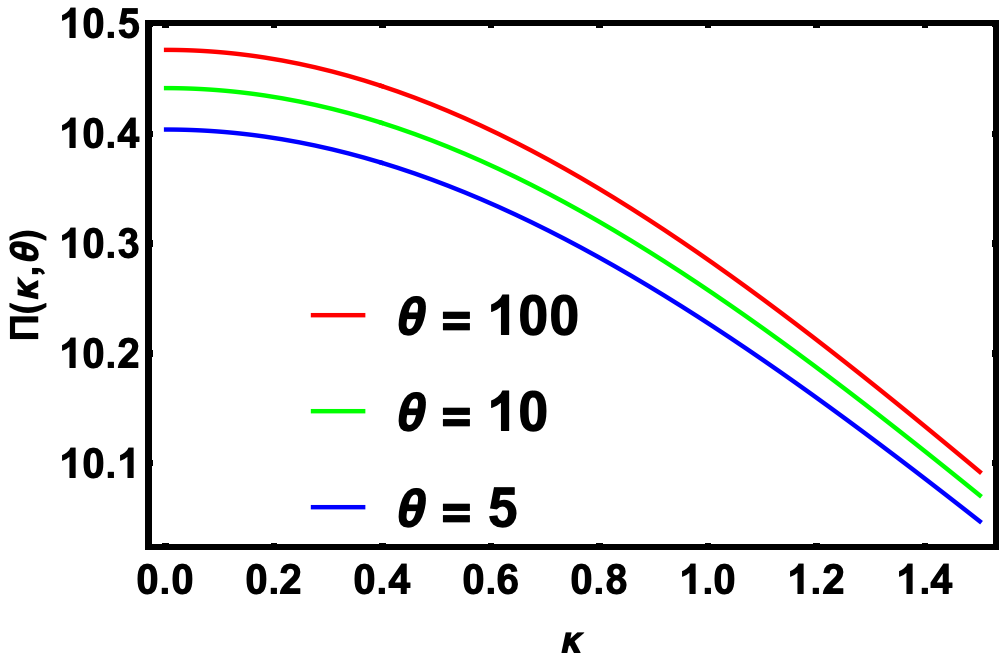}
\includegraphics[width=3in,height=2in]{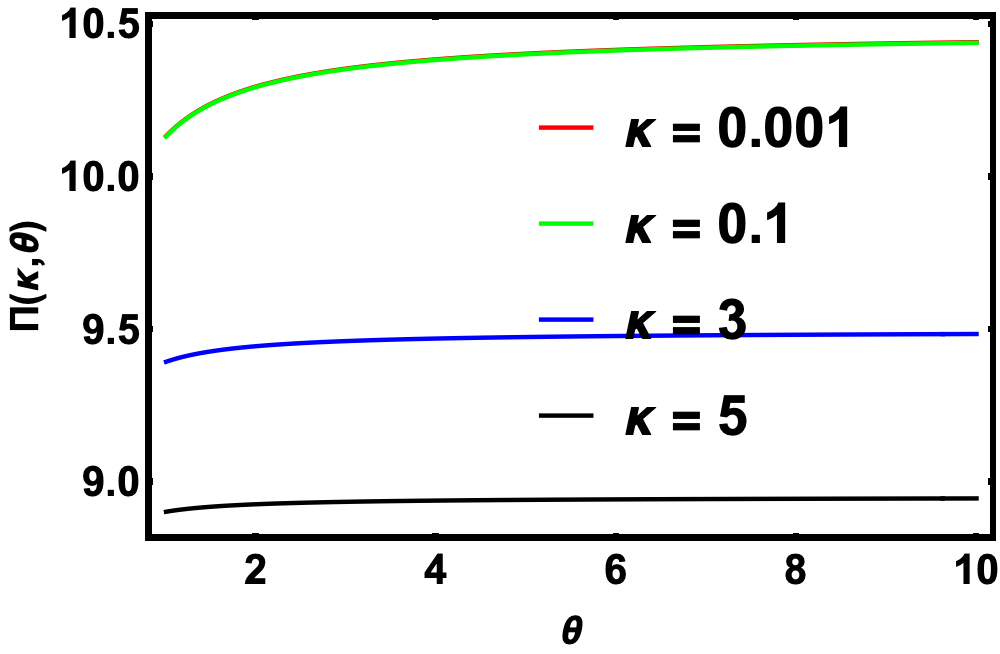}
\caption{Same caption as Fig.~\ref{LargeKLargeT} but for $\kappa \ll 1$ and $\theta \gg 1$. These results demonstrate that in the limit $\kappa\ll1$ and $\theta \gg1$, impurity vertex corrections have a discernible yet small effect on the pair susceptibility.  } \label{SmallK}
\end{figure}
\begin{figure}[h!]
\includegraphics[width=3in,height=2in]{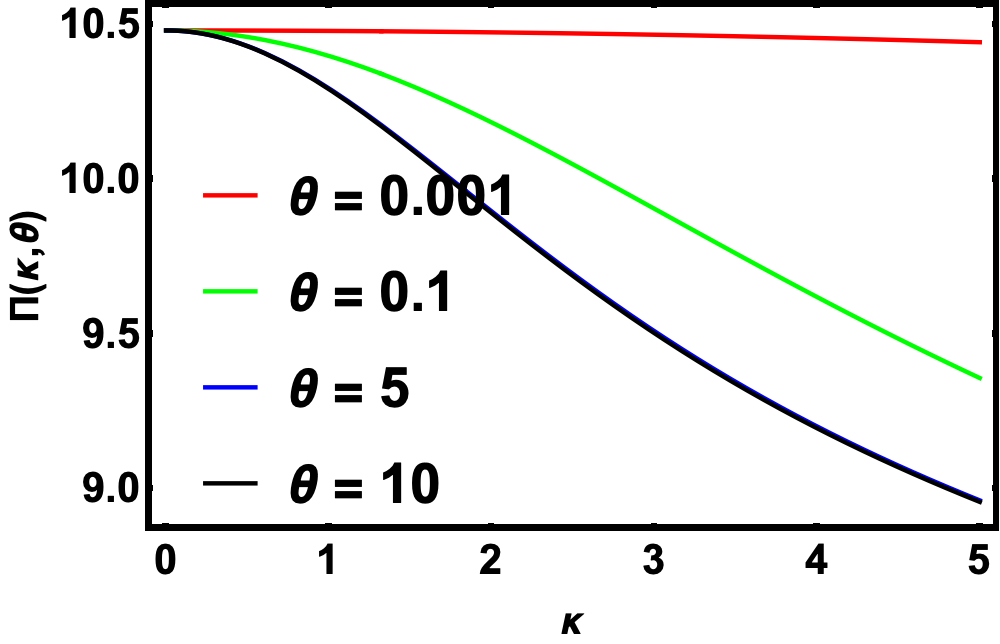}
\includegraphics[width=3in,height=2in]{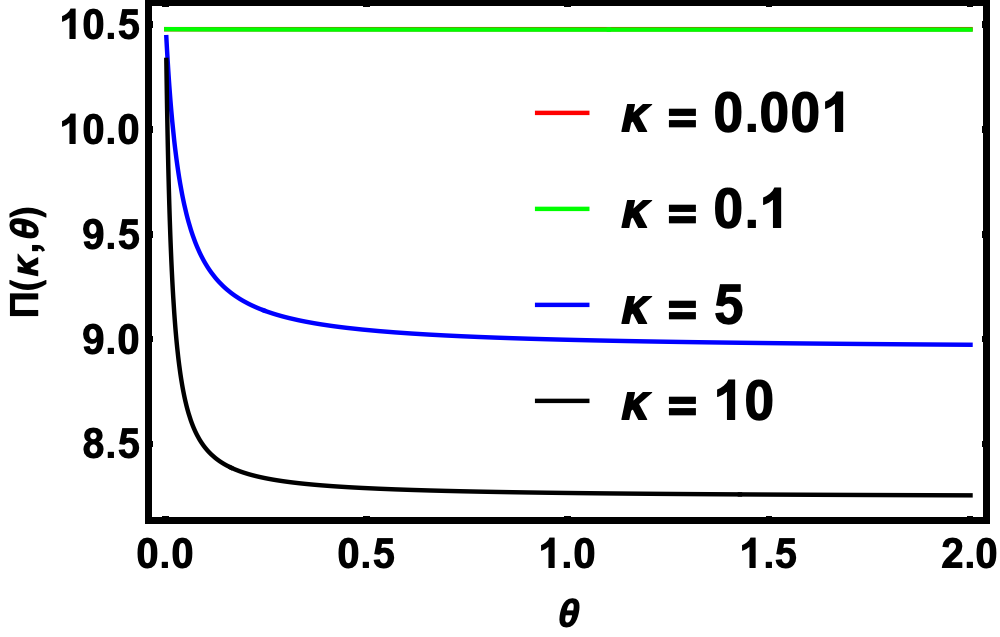}
\includegraphics[width=3in,height=2in]{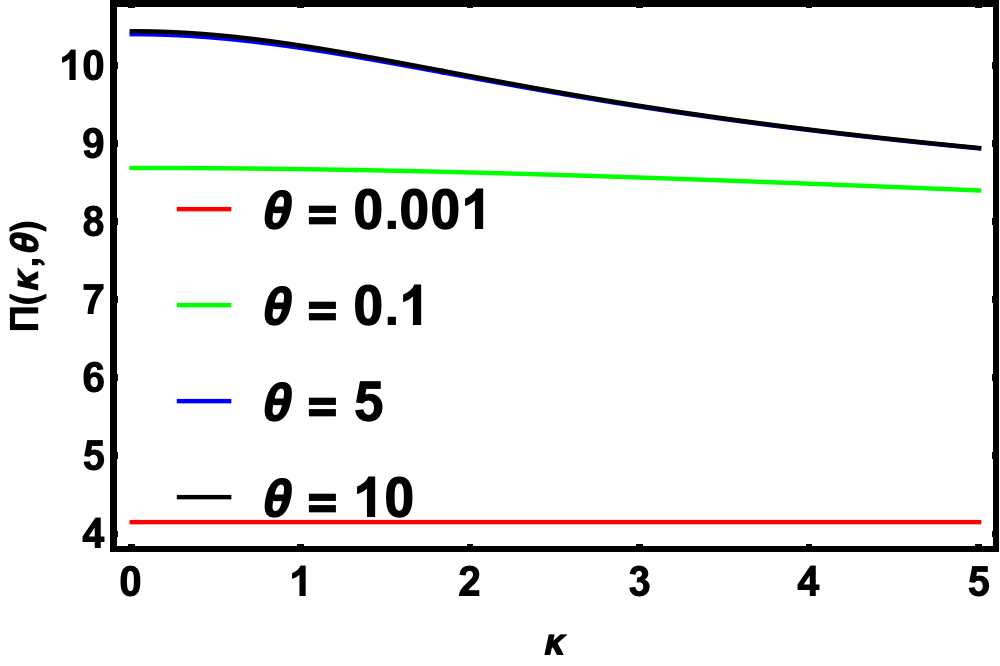}
\includegraphics[width=3in,height=2in]{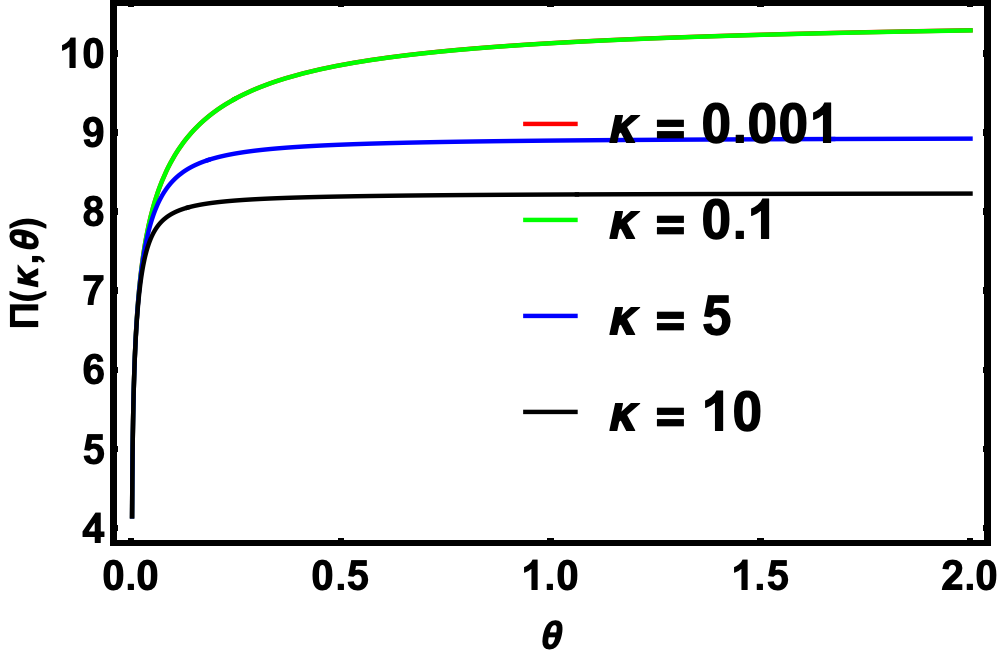}
\caption{Same caption as Fig.~\ref{LargeKLargeT} but for $\theta \ll 1$ over a range of $\kappa$. These results demonstrate that in the limit $\theta \ll1$, impurity vertex corrections completely alter the pair susceptibility leading to an eventual breakdown of perturbation theory. } \label{SmallK}
\end{figure}
\newpage
\textcolor{black}{
\section{Appendix H}
In this appendix we derive the form of the free-energy in Eq.~\ref{FreeEnergy} from standard methods employed in Refs.~\cite{LarkinVarlamov2005, Vinokur2005}. A crucial step in studying the fluctuation properties of a generic multi-band superconductor involves calculation of the fluctuation propagator $L^{-1}_{\mu\nu}(\bs q, \Omega) = -i \Omega \gamma_{\mu \nu}(\bs q) + H_{\mu \nu}(\bs q)$. Here $\mu$ and $\nu$ are internal indices like band/orbital, $\gamma_{\mu\nu}$ are dynamic coefficients and $H_{\mu\nu}$ is the Hamiltonian density. The fluctuation propagator is a matrix coefficient of the quadratic (Gaussian) term in the Ginzburg-Landau free-energy expansion. In our case, the fluctuation propagator has been obtained phenomenologically from the YRZ Green function.  With knowledge of the Hamiltonian density, the fluctuation partition function to Gaussian order is defined by
\beq
Z = \int\mathscr{D}\Delta\int\mathscr{D}\Delta^* exp\left[ -\beta \int \frac{d^d \bs q}{(2\pi)^d}\Delta^*H(\bs q) \Delta \right].
\eeq
 We can now write the fluctuation contribution to the free energy as a functional integral of a gaussian for the simple one-band ($\mu = \nu=1$) case in $d-$dimensions as
\beq
F &=& -T~log~Z \\
 &=& -T~log\int\mathscr{D}\Delta\int\mathscr{D}\Delta^* exp\left[ -\beta \int \frac{d^d \bs q}{(2\pi)^d}\Delta^*H(\bs q) \Delta \right]. \\
&=&  T V \int\frac{d^d \bs q}{(2\pi)^d} ~log \left(H(\bs q)\right) + c,
\eeq
where $V$ is the volume and $c$ is a constant independent of temperature. To obtain $H(\bs q)$, we can use the static limit of the fluctuation propagator; from the equation above, this gives $H(\bs q)=L^{-1}(\bs q, \Omega=0) $. In the case of interest, we have from Eq.~\ref{NonCriticalPropagator} of the main text 
\beq \nonumber
L^{-1}(\bs q, \Omega = 0) &\simeq& - g^{-1} + N_0 \left[\ln\frac{\Lambda}{u} + \sqrt{\frac{\pi \kappa}{2}} e^{-\kappa}\right] - \frac{N_0 r^2}{12 u^2} \left[ 1 + \sqrt{\frac{\pi \kappa^3}{8}} e^{-\kappa} \right].
\eeq
 In principle the entire $\bs q$ integration can be performed. However, as we are interested in the limit where the spatial structure is unimportant so as to make a comparison with a zero dimensional SYK model, we only need to consider contribution to the integral from the $\bs q \rightarrow 0$ mode. This limit of the fluctuation propagator has been written in Eq.~\ref{CriticalPropagator} of the main text where $g$ has been substituted in favor of $u_{c\infty}$ close to the QCP, and is given by 
 \beq 
L^{-1}(\bs q\rightarrow0, \Omega = 0)_{u=u_{c\infty}} = N_0 \sqrt{\frac{\pi u_{c\infty}}{2 T}} e^{\frac{-u_{c\infty}}{T}}.
\eeq
  Substituting for $L^{-1}(\bs q\rightarrow0, \Omega = 0)_{u=u_{c\infty}}$ into the Hamiltonian density, the free energy intergral can now be performed trivially where the momentum space volume cancels the volume factor $V$. Hence, along with an additional constant independent of temperature, one is left with Eq 8 of the manuscript given as
   \beq
 -\beta F = \beta u_{c\infty} - \gamma~\ln(\beta u_{c\infty}),
 \eeq  
 where $\gamma = 1/2$. }
\end{document}